\documentclass[twocolumn]{aastex631}

\usepackage{float}
\usepackage{comment}
\usepackage{xcolor}
\usepackage{lipsum}
\usepackage{enumitem}

\usepackage{xspace}

\newcommand{\HI}{\ion{H}{1}\xspace}
\newcommand{\HII}{\ion{H}{2}\xspace}
\newcommand{\kms}{\ensuremath{\textrm{ km s}^{-1}}\xspace}
\newcommand{\Ha}{\ensuremath{\textrm{H}\alpha}\xspace}

\defcitealias{Arvidsson2009}{A09}
\defcitealias{Booth_2022}{B22}
\defcitealias{basu1999dynamical}{B99}
\defcitealias{west2007fragmenting}{W07}
\defcitealias{stil2009three}{S09}
\defcitealias{Kompaneets1960}{Kompaneets model}
\defcitealias{Ziegenbalg2025}{Ziegenbalg}

\begin{document}

\title{G92.4+5.0: A Perseus Arm superbubble associated with the \HII region CTB~102}

\newcommand{\UBCO}{Department of Math, Physics, \& Statistics, University of British Columbia, Okanagan Campus, Kelowna, BC, V1V 1V7, Canada}

\newcommand{\UCalgary}{Department of Physics and Astronomy, University of Calgary, 2500 University Drive NW, Calgary, Alberta, T2N 1N4, Canada}

\newcommand{\DRAO}{Dominion Radio Astrophysical Observatory, Herzberg Research Centre for Astronomy and Astrophysics, National Research Council Canada, PO Box 248, Penticton, BC, V2A 6J9, Canada}

\newcommand{\UWO}{Department of Physics \& Astronomy, University of Western Ontario, 1151 Richmond Street, London, ON, N6A 3K7, Canada}

\newcommand{\UofA}{Department of Physics, University of Alberta, 11322 -- 89 Avenue, Edmonton, Alberta, T6G 2E1, Canada}

\newcommand{\Brandon}{Department of Physics and Astronomy, Brandon University, 270-18th Street, Brandon, MB, Canada}

%UBCO, UCalgary, DRAO, UWO, UofA, Brandon
%UBCO, UCalgary, DRAO, UWO, UofA, Brandon

\author[0000-0001-5181-6673]{Rebecca A. Booth}
\affiliation{\UBCO}
\affiliation{\UCalgary}
\affiliation{\DRAO}

\author[0000-0003-4781-5701]{Jo-Anne Brown}
\affiliation{\UCalgary}

\author[0000-0003-1455-2546]{Tom Landecker}
\affiliation{\DRAO}
\affiliation{\UCalgary}

\author[0000-0001-5953-0100]{Roland Kothes}
\affiliation{\DRAO}
\affiliation{\UofA}

\author[0000-0002-3189-4882]{Tyler Foster}
\affiliation{\Brandon}

\author[0000-0002-2465-8937]{Anna Ordog}
\affiliation{\UWO}
\affiliation{\UBCO}
\affiliation{\DRAO}

\author[0000-0001-7301-5666]{Alex Schwaner Hill}
\affiliation{\UBCO}
\affiliation{\DRAO}

\begin{abstract}

Superbubbles are vast hot cavities carved out by the stellar winds and supernovae of clusters of high-mass stars. We investigate G92.4$+$5.0 (the CTB~102 superbubble), a Perseus Arm superbubble associated with the \HII region CTB 102. We use new and archival radio observations (total intensity, rotation measure, and \HI spectral line) from the Dominion Radio Astrophysical Observatory Synthesis Telescope, combined with single-antenna data from the Effelsberg Medium Latitude Survey, the Global Magneto-Ionic Medium Survey, and the HI4PI survey to provide data on large structures. The upper region of the superbubble is faint in \HI. In order to detect its full extent, we supplement our observations with \Ha data from the Wisconsin H-Alpha Mapper Survey and the Ziegenbalg Survey, and observe \Ha emission in the upper region. We measure the CTB~102 superbubble to be at least $\sim170$ pc wide and $\sim420$ pc tall, with an expansion rate of $\sim12$~\kms in the lower regions of the superbubble and $\sim22.5$~\kms in the upper regions. The elongated structure and increasing expansion rate with height are consistent with the evolutionary properties of a superbubble described by the Kompaneets model. We fit the Kompaneets model to the CTB~102 superbubble and obtain an estimated age of 1--6 Myr. This supports a possible scenario in which the stellar cluster currently ionizing CTB 102 is the expansion energy source for the superbubble. Alternatively, if the superbubble expanded under the power of a previous generation of stars, the current CTB~102 stellar cluster may have formed in the walls of the superbubble.

\end{abstract}


\section{Introduction} \label{sec:intro}

The stellar winds and supernovae of high mass stars leave a lasting imprint on the interstellar medium (ISM) through the formation of bubbles, which are large, hot cavities bounded by walls of swept-up gas and magnetic fields. In turn, the expanding shock front and compressed fields at the bubble walls trigger new star formation \citep[e.g.,][]{Oey2005, Zucker2022}, and the cycle repeats. The largest bubbles are formed by the combined winds and supernovae of a group of stars. These `superbubbles' (SB) reach radii of hundreds of pc and are major features of the ISM. 

During early stages of their evolution, SBs tend to be approximately spherical in shape, expanding at the same rate in all directions \citep{basu1999dynamical}. As they age, SBs become more elliptical, expanding rapidly in the vertical direction relative to the Galactic plane due to lower mechanical pressure as density decreases with height. Furthermore, SBs become elongated in the direction of the ambient magnetic field, as magnetic pressure resists expansion perpendicular to the field \citep{Tomisaka1998, stil2009three}. During the late stages of SB evolution, the SB walls fragment and disperse, enriching the ISM with energy and the products of stellar fusion. In the case where a bubble breaks out of the Galactic disk before fragmentation, hot matter is funneled into the halo as the bubble becomes a chimney \citep{mac1988numerical, norman1989disk}.

Many models have been developed to understand the evolution of bubbles, ranging from a purely analytical framework \citep[e.g.,][]{Kompaneets1960, Castor1975ApJ, Weaver1977ApJ} to magneto-hydrodynamic simulations covering timescales of hundreds of millions of years \citep[e.g.,][]{Tomisaka1998, deAvillez2005, stil2009three, hill2012vertical}. Testing these models requires observations of many SBs at various stages of evolution. 

In this paper, we examine observations of a SB associated with CTB~102, an \HII region located at a distance of 4.3~kpc in the Perseus Spiral Arm (\citeauthor{Arvidsson2009} \citeyear{Arvidsson2009}; hereafter \citetalias{Arvidsson2009}). CTB~102 is primarily a radio source, first recorded in the California Institute of Technology 960~MHz B catalog by \citet{wilson1960survey}. At a distance of 4.3~kpc, optical extinction effects are significant, and the stellar cluster ionizing CTB~102 has not been observed. However, \citetalias{Arvidsson2009} estimated that a minimum of $4.5\times10^{49}$ ionizing photons per second power CTB~102, requiring at least several late type O stars or one early type O star to be present in the region. CTB~102 has also been shown to be associated with molecular clouds supporting significant star formation activity across a large region with dimensions of 60 $\times$ 35 pc$^2$ \citep{marshall2019high}. 

In their analysis of a series of \HI filaments associated with CTB~102, \citetalias{Arvidsson2009} suggested the \HII region may be embedded in a large bubble or chimney, carved out by the stellar winds of the same star(s) powering the \HII region. In this work, we revisit this potential bubble associated with CTB~102 on the basis of archival radio continuum, \HI spectral line, and \Ha observations. We show that while the morphology of the \HI filaments around CTB~102 appears to open like a chimney, the inclusion of \Ha observations reveals a cap. In Section \ref{sec:data}, we describe the datasets we use, and in Section \ref{sec:results}, we present the results. In Section \ref{sec:analysis}, we analyze the results by applying the model of \citeauthor{Kompaneets1960} (\citeyear{Kompaneets1960}; hereafter \citetalias{Kompaneets1960}) to observations of the bubble morphology. In Section \ref{sec:discussion}, we discuss the several possible origin scenarios for the bubble.

%%%%%%%%%%%%%%%%%%%%%%%%%%%%%%%%%%%%%%%%%%%%%%%%%%%%%%%%%%%%%%%%%%%%%%%%%%%%%%%%%%%%

\section{Observations} \label{sec:data}

We use observations covering Galactic longitudes $88.9^\circ < \ell <  99.1^\circ$ and latitudes $0.1^\circ < b < 10.3^\circ$. Our data include archival continuum and \ion{H}{1} spectral line observations by the Dominion Radio Astrophysical Observatory Synthesis Telescope \citep[DRAO-ST;][]{landecker2000synthesis}. Additionally, we make use of two published H$\alpha$ datasets, the Wisconsin H-Alpha Mapper (WHAM) survey \citep{haffner2003wisconsin}, and the Northern Sky Narrowband survey (\citeauthor{Ziegenbalg2025} \citeyear{Ziegenbalg2025}; hereafter the Ziegenbalg survey). We describe these data below. 

\subsection{DRAO-ST observations}

\begin{table}[t!]
\centering
\begin{center}
\begin{tabular}{ll}
\hline
Coverage     &  ${88.9^{\circ}} < {\ell} < {99.1^{\circ}}$\\
                                  & ${0.1^{\circ}} < {b} < {10.3^{\circ}}$\\

Angular resolution           & $58'' \times 58''$\,cosec\thinspace$\delta$ \\
Sensitivity, $I$             & 0.30\,mJy/beam rms \\
%Sensitivity, $Q$ and $U$     & 0.25\,mJy/beam rms \\
Typical noise in mosaicked   & ${76}\thinspace{\rm{sin{\thinspace}}\delta}$~mK \\
~~~~~~~~~~~~~~~~~~~~~~~~~~images   & \\

Sources of single-antenna data & Effelsberg 100-m Telescope \\
                             & and DRAO 26-m Telescope \\
%\hline
%\textbf{Continuum observations} & \\
Continuum Bandwidth          & 30\,MHz in four bands of \\
                             & ~~~~~~~~~~~~~~7.5\,MHz each \\
Band center frequencies      & 1407.2, 1414.1, 1427.7,\\
                             & ~~~~~~~~~and 1434.6\,MHz \\
Polarization products        & Stokes $I$, $Q$, and $U$ \\
%\hline           
%\textbf{\ion{H}{1} data cube} & \\
\ion{H}{1} velocity coverage & $-$163.9 to 46.3~\kms \\
Velocity resolution          & 1.3 km~s$^{-1}$ \\
\ion{H}{1} sensitivity       & 2.2~K rms in a single channel \\
%(mosaicked data)             & \\
\hline
\end{tabular}
\caption[]{Properties of the DRAO-ST Observations}
\label{observing_specs}
\end{center}
\end{table}

The DRAO-ST data comprise total intensity (Stokes $I$) and linear polarization (Stokes $Q$ and $U$) maps of continuum emission in four frequency channels, 1407.2, 1414.1, 1427.7, and 1434.6~MHz, and an \ion{H}{1} spectral line data cube covering the velocity range $-163.9$ to 46.3 km~s$^{-1}$, with a spectral resolution of 1.3~\kms \citep{landecker2000synthesis}. We summarize the properties of these data in Table \ref{observing_specs}. Latitudes up to $b=5.6^\circ$ were observed as part of the Canadian Galactic Plane Survey \citep[CGPS;][]{tayl03}. % and have been previously discussed in \citetalias{Arvidsson2009}. 
The data 
%in the upper Galactic region, 
spanning latitudes from 5.6$^\circ$ to 10.4$^\circ$ were obtained in a later 2012 observation. We mosaicked the CGPS data and 2012 observation using the Supertile tool, which is included in the DRAO Export Package \citep{Higgs1997}. 

% NOTE: CHECK \citep{land10}
At 1420~MHz, the DRAO-ST provides an angular resolution of $\sim58^{\prime\prime}$ for total power continuum emission and $\sim58^{\prime\prime}$ for \ion{H}{1} and polarization. The smallest antenna baseline of 12.86~m allows for sensitivity to angular structures as large as 0.5$^\circ$ \citep{landecker2000synthesis}. To ensure sensitivity to large-scale structure, we combined the Synthesis Telescope data with single-antenna observations. We combined the total intensity continuum data with the Effelsberg Medium Latitude Survey \citep[EMLS;][]{reic04} and the \HI data with the HI4PI survey \citep{bekhti2016hi4pi} using the method described by \citet{tayl03}. We combined the polarization observations with data from the Global Magneto-Ionic Medium Survey \citep{wolleben2021global}, observed with the DRAO 26-m Telescope, using the method described by \citet{ordog2025pioneering}.

\subsection{Rotation measure observations}

Polarized emission undergoes Faraday rotation as it traverses a path containing a magnetic field and free electrons. As a result, the polarization angle, $\tau$, %= 1/2 \arctan(U/Q)$, 
rotates depending on the Rotation Measure (RM) and wavelength squared ($\lambda^2$) as $\Delta\tau =\textrm{RM}\,  \lambda^2$. The RM of a line-of-sight (LOS) is defined as the path integral, 
\begin{equation}
\frac{\textrm{RM}}{\textrm{rad/m}^2} = 0.812 \int_{\textrm{source}}^{\textrm{observer}} \bigg(\frac{n_e}{\textrm{cm}^{-3}} \bigg) \bigg(\frac{B_\parallel}{\mu\textrm{G}} \bigg)\, \bigg(\frac{dr}{\textrm{pc}}\bigg).
\end{equation}
where $r$ is the distance along the path, $n_e$ is electron density, and $B_\parallel$ is the LOS component of the magnetic field.

%%%%%%%%%%%%%%%%%% Continuum map %%%%%%%%%%%%%%%%%%%%%%%%
\begin{figure}[b!]
    \includegraphics[width=0.47\textwidth]{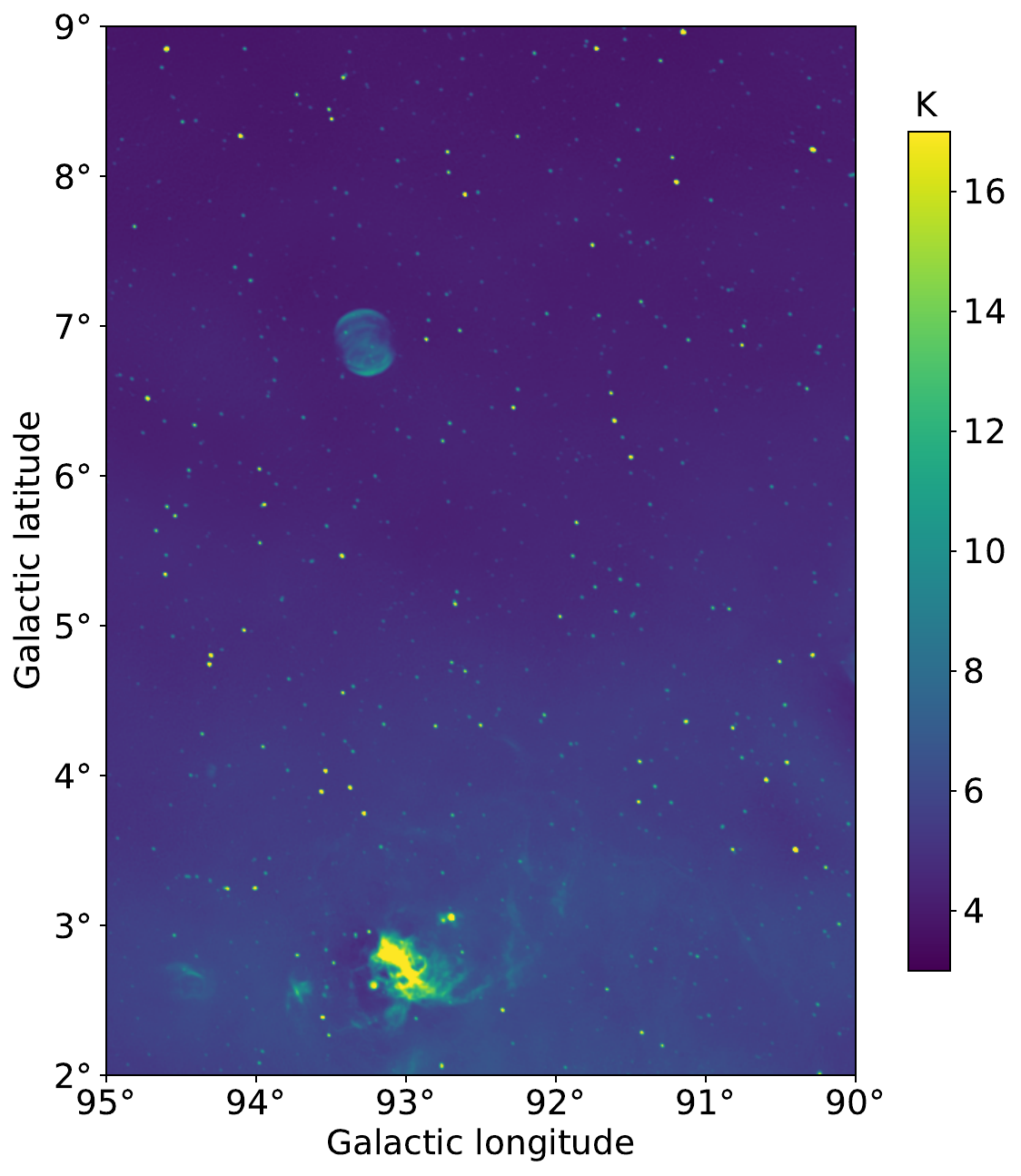}\\
    \caption{Total intensity continuum emission (Stokes I) at 1420~MHz from combined DRAO-ST and EMLS observations. The \ion{H}{2} region CTB~102 is located at $(\ell, b) = (93^\circ, 2.8^\circ)$. The supernova remnant DA~530 is located above at $(\ell, b)~=~(93.3^\circ, 6.9^\circ)$.}
    \label{fig:continuum_and_P}
\end{figure}
%%%%%%%%%%%%%%%%%%%%%%%%%%%%%%%%%%%%%%%%%%%%%%%%%%%%%%%%%

For this work, we have prepared an RM map %comprising slopes determined from linear fits to $\tau$ vs $\lambda^2$ 
using our combined DRAO-ST and GMIMS polarization data across the four continuum frequency channels. With so few frequency channels, the magnitude and even the signs of the RMs in our map are likely not reliable if there is any Faraday complexity \citep{hill2018there, ordog2025pioneering}. Appreciating this limitation, these RM data still %we can still extract 
offer valuable information in the form of spatially coherent RM structures. These indicate changes in Faraday rotation conditions as the LOS sweeps across a region and provide evidence of electron density and magnetic field enhancements.

\subsection{H$\alpha$ observations}

We supplement our DRAO-ST observations with published H$\alpha$ data products from the WHAM Northern Sky Survey and Ziegenbalg Survey. The WHAM data comprise absolutely calibrated H$\alpha$ spectra, with 1$^\circ$ angular resolution, 12 km~s$^{-1}$ spectral resolution \citep{haffner2003wisconsin}, and a sensitivity of $\sim$0.15 R. 

Due to WHAM's limited angular resolution, we use the Ziegenbalg H$\alpha$ data to supply fine detail. The Ziegenbalg Survey is an `amateur'-led astronomy project observed using private instruments in Pulsnitz, Germany \citep{Ziegenbalg2025}. The Ziegenbalg H$\alpha$ data have a resolution of 9$^{\prime\prime}$, are approximately calibrated in Rayleighs, and cover the entire sky down to a decl. of $-16^\circ$. These data were observed with 4892 hours of exposure time, and, with a bandpass of 3.5~nm, they contain the entire LOS integrated into a single map. 

%%%%%%%%%%%%%%%%%%%%%%%%%%%%%%%%%%%%%%%%%%%%%%%%%%%%%%%%%%%%%%%%%%%%%%%%%%%%%%%%%%%%

\section{Results} \label{sec:results}

Figure \ref{fig:continuum_and_P} displays the DRAO-ST total intensity continuum map averaged through the four continuum frequency channels. In this map, the \ion{H}{2} region CTB~102, centered at $(\ell, b) = (93^\circ, 2.8^\circ)$, is the most prominent continuum emission source. The supernova remnant (SNR) DA~530 appears above, located at $(\ell, b) = (93.3^\circ, 6.9^\circ)$. There are several faint filamentary structures curving up and to the right from the brightest region of CTB~102. \citetalias{Arvidsson2009} postulated that these are the ionized components of a cavity formed by stellar activity within the CTB~102 complex. Hereafter we refer to this cavity as ``the CTB~102 SB'' (or just ``the SB''). In the following subsections, we revisit the SB structures identified by \citetalias{Arvidsson2009} in our DRAO-ST RM map and extended latitude \ion{H}{1} data, and present new insights provided by the H$\alpha$ data.

\subsection{Rotation measure results} \label{sec:RMobs}

\begin{figure}[t!]
    \includegraphics[width=0.49\textwidth]{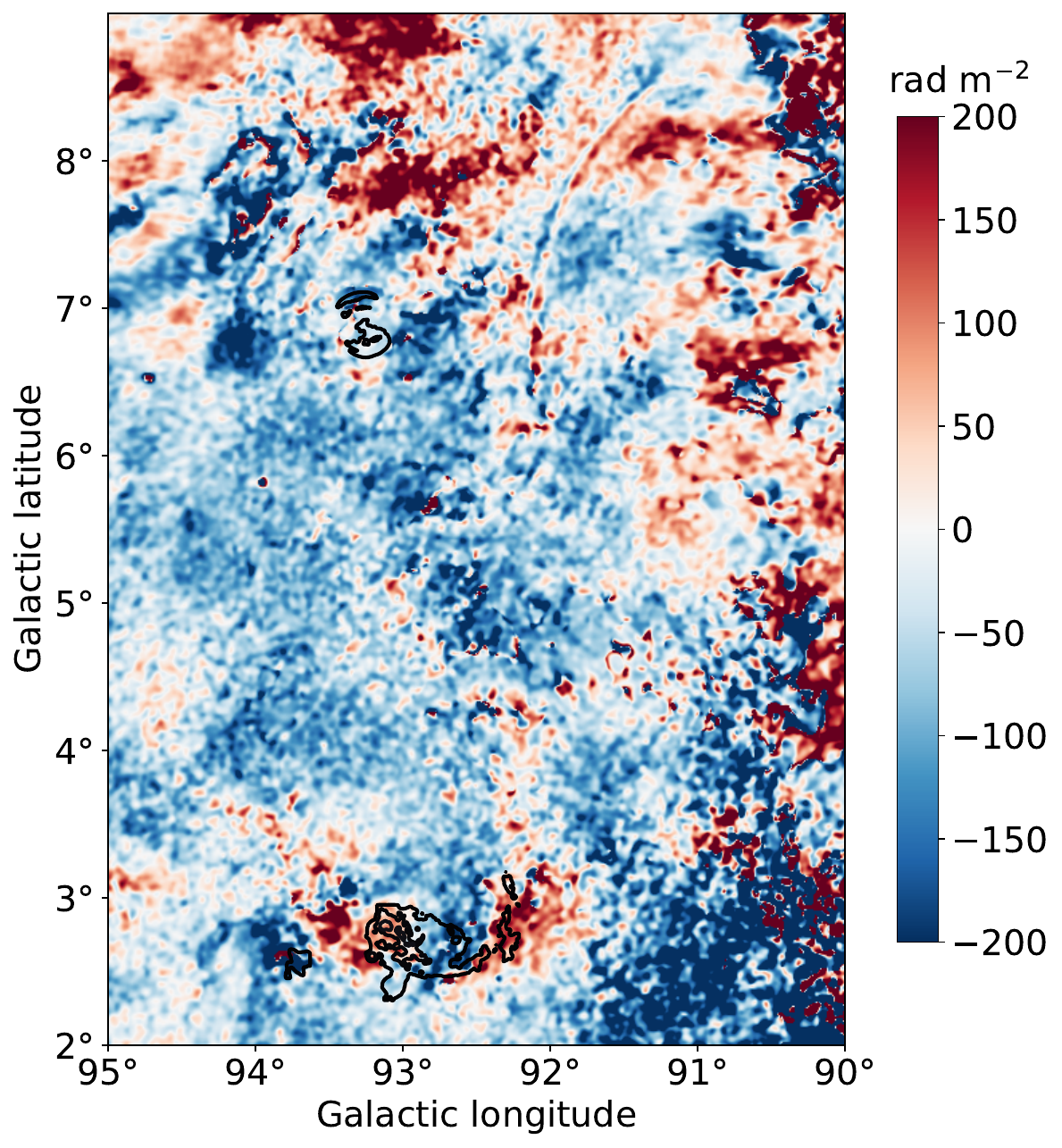}\\
    \caption{Rotation measure map produced using combined DRAO-ST and GMIMS polarization observations at the four DRAO-ST continuum frequency channels, 1407.2, 1414.1, 1427.7, and 1434.6~MHz. Continuum emission contours from Figure \ref{fig:continuum_and_P} are included around CTB~102 and DA~530 at 7K, 14K, and 28 K.}
    \label{fig:RMfigure}
\end{figure}

Figure \ref{fig:RMfigure} shows the RM map. %, with continuum contours for CTB~102 and DA~530 included for reference. 
In this figure, a region of positive RMs forms a `U-shaped' bowl surrounding CTB~102. This U-shaped RM structure is well defined against the generally negative background and has a clear boundary, evidence that the Faraday rotation along these sightlines is dominated by magnetic fields and ionization from a discrete object or region. The morphological agreement of the RM structure with the curved filaments of CTB~102 suggests that the Faraday rotation is associated with magnetic field and electron density enhancements in the vicinity of the \HII region. This supports the idea of \citetalias{Arvidsson2009} that CTB~102 sits within a larger bubble/chimney complex. In this case, an enhanced magnetic field could result from stellar magnetic fields carried by stellar winds \citep[e.g.,][]{harv10} or the compression of the ambient magnetic field by the expanding shock at the SB walls \citep{Tomisaka1990}. Electron density enhancements could be supplied by shock ionization at the SB walls or photoionization by photons emitted from the stellar cluster that powers CTB~102.

%%%%%%%%%%%%%%%%%%%%%%%% RM map %%%%%%%%%%%%%%%%%%%%%%%%

\subsection{\HI spectral line results reveal the superbubble}

\begin{figure*}[ht!]
  \centering
    \includegraphics[width=0.98\textwidth]{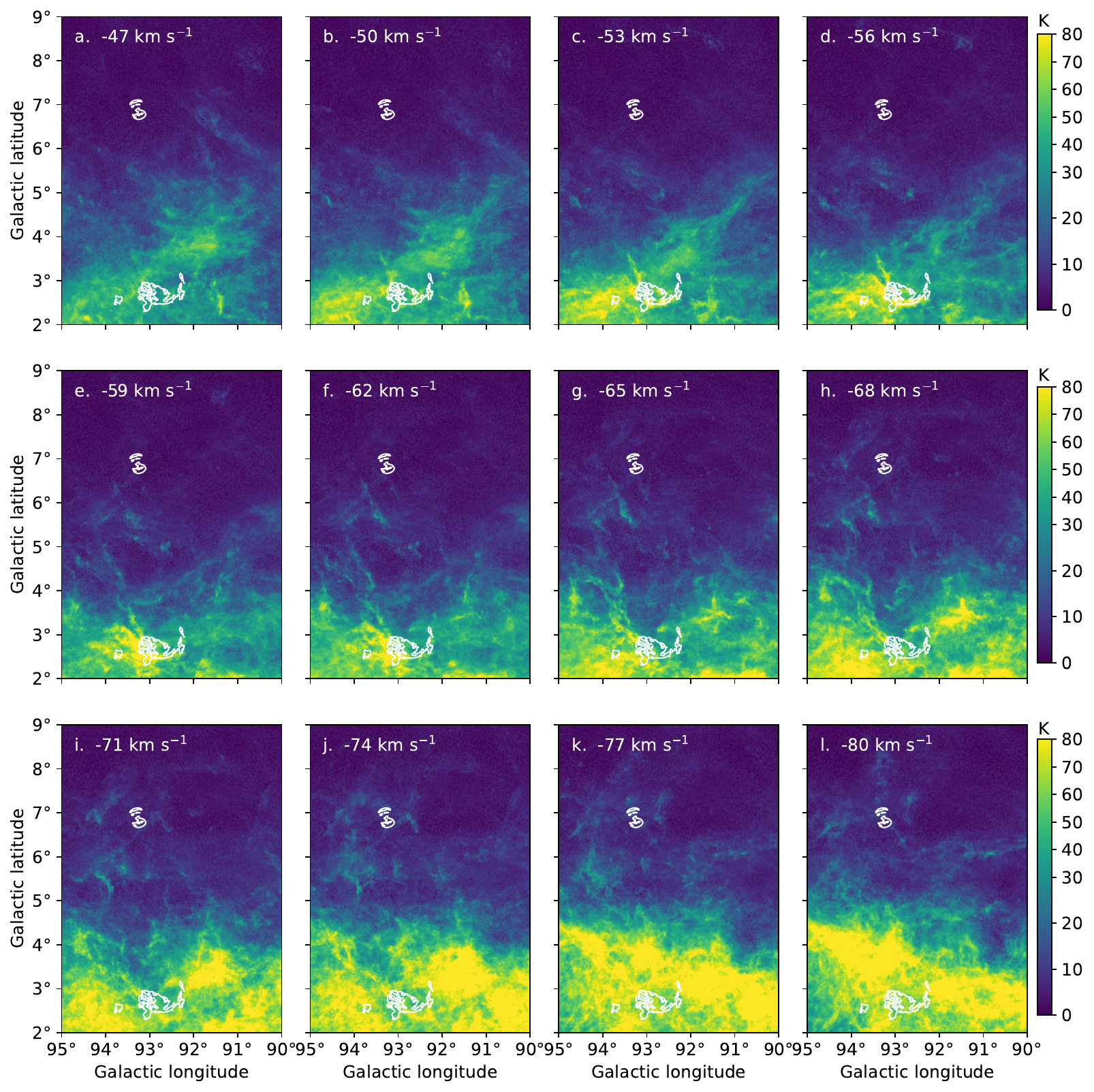}
    \caption{\HI spectral images from the combined DRAO-ST and HI4PI data cube. Continuum emission contours from Figure \ref{fig:continuum_and_P} are included around CTB~102 and DA~530 at 7K, 14K, and 28 K.}
    \label{fig:HIa}
\end{figure*}

In Figure \ref{fig:HIa}, we display selected images from our \HI data cube through the velocity\footnote{All velocities in this paper are references to the local standard of rest.} range $-80<v<-47$~\kms. \citetalias{Arvidsson2009} identified \HI filaments within this velocity range that they interpreted to be the walls of the SB. We observe the same filaments in Figure \ref{fig:HIa}, with a structure that appears to wrap around the base of CTB 102 starting from $v=-50$~\kms (panel b) to $v=-62$~\kms (panel f). The structure resembles a backwards `J-shape', with the higher left side mirroring the curved continuum filaments on the right side, and continues through velocity channels $-60 \leq v \leq -50$~\kms. 

In the more negative velocity range of $-61 \leq v \leq -74$~\kms, the \HI structures surrounding CTB~102 appear as a `V-shaped' cavity. Observing this V-shape, \citetalias{Arvidsson2009} postulated that the SB may be an open chimney. With the higher latitudes included in our new \HI data, we continue to observe no evidence of a closure at the top of the structure and our \HI data appear to support this chimney hypothesis. %At around $v\sim-74$\kms, the V-shaped structure starts to merge with more widespread Galactic \HI emission, and all hints of this structure vanish beyond $-77$~\kms.

\subsubsection{The near and far walls of the superbubble} \label{HIsidewalls}

\begin{figure*}[ht!]
\centering
    \includegraphics[width=1\textwidth]{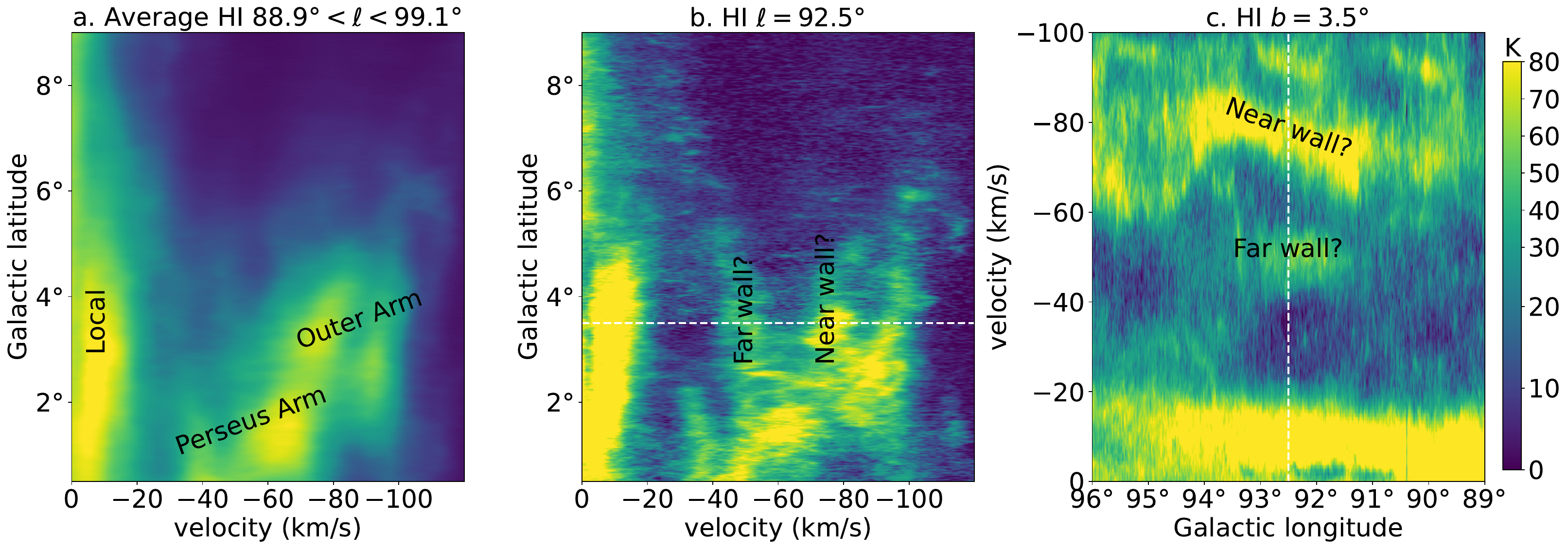}
    \caption{Alternative views through the \HI data cube. \textit{a.} A velocity--latitude plot averaged through all Galactic longitudes in the \HI data cube. Widespread emission due to Galactic structure is identified. \textit{b.} A velocity--latitude plot of the \HI cube data through longitude $\ell = 92.5^\circ.$ Emission due to the potential near and far walls of the superbubble is identified. \textit{c.} A longitude--velocity plot through latitude $b=3.5^\circ$ with the potential near and far superbubble walls identified. The white dashed line in \textit{b} (\textit{c}) indicates the latitude (longitude) of \textit{c} (\textit{b}). The velocity axis increases through negative velocities because for Galactic structure more-negative velocities correspond to larger distances in the sightlines of our observations.}
    \label{fig:HIb}
\end{figure*}

Figure \ref{fig:HIb} shows alternative views through the \HI data cube. In panel a, we averaged the cube through all Galactic longitudes and display the resulting velocity--latitude ($v$--$b$) image. This image enables the identification of widespread emission structures associated with the Galactic spiral arms and helps us differentiate between Galactic emission and potential SB structure. In the $v$--$b$ image, Perseus Arm emission can be observed between  $-75 \leq v \leq -35$~\kms, sloping upwards in latitude at more negative velocities due to the ``rolling motions'' in the arm \citepalias{Arvidsson2009}. Around $v\sim-70$~\kms, Outer Arm emission blends with the emission from the Perseus Arm. The Outer Arm emission appears at higher latitudes due to the Galactic warp, and in the velocity range  $-100 \leq v \leq -70$~\kms.

Figure \ref{fig:HIb}b shows a $v$--$b$ slice of the \HI data cube through the centre of the V-shape, at longitude $\ell = 92.5^\circ$. In this plot, we observe two bands of \HI emission at $v\sim -50$ and $\sim-75$~\kms. These may represent the near and far walls of the SB, with the near wall at the more negative velocity ($v\sim -75$~\kms) as it is expanding towards us and the far wall at the less negative velocity ($\sim -50$~\kms) as it is expanding away from us. The potential far wall rises above the Perseus arm emission, which tends to be at lower Galactic latitudes around $v\sim -50$~\kms. This makes the far wall easy to distinguish from Galactic emission. The potential near wall, however, coincides with the region in velocity space where the emission from the Perseus Arm and Outer Arm blend into each other and is less certain.

%%%%%%%%%%%%%%%%%%%%%%%% HI cube %%%%%%%%%%%%%%%%%%%%%%%%%
%%%%%%%%%%%%%%%%%%%%%%%%%%%%%%%%%%%%%%%%%%%%%%%%%%%%%%%%%%

%%%%%%%%%%%%%%%%%%%%%%%% Other HI figure %%%%%%%%%%%%%%%%%%%

%%%%%%%%%%%%%%%%%%%%%%%%%%%%%%%%%%%%%%%%%%%%%%%%%%%%%%%%%%%%

We observe the same two potential SB walls of emission in Figure \ref{fig:HIb}c, which shows a longitude--velocity slice of the \HI data cube. The slice is cut along $b=3.5^\circ$, the latitude that passes through both the stem of the J-shape and the low-emission region of the V-shape. Here, a narrow vertical line at $\ell = 93^\circ$ extends from  $-70 \leq v \leq -50$~\kms and corresponds with the stem of the J-shape and the left side of the V-shape. We take this to be a side wall of the SB. The candidate far wall appears as a horizontal band of emission along $v\sim-50$~\kms. Three pieces of evidence link this band to the SB: it remains relatively isolated from other emission sources; it covers the observed longitude range of the SB; it appears to connect to the side wall. We are therefore confident that it is likely the far wall. The potential near wall appears as a second band at $v \sim -75$~\kms. As this band blends with the broader Galactic emission extending across the entire longitude range, we continue to be uncertain whether it is a SB wall, Galactic structure, or both.

\subsection{H$\alpha$ results reveal a cap} \label{sec:Ha}

We have identified that the \HI\ morphology around CTB~102 resembles an open chimney, consistent with the observations of \citetalias{Arvidsson2009}. However, the neutral gas component alone may not tell the full story, as any ionized components of the SB walls would be missing in \HI data. We therefore turned to \Ha emission to look for evidence of a cap at the top of the SB. There is precedent for this approach: the full extents of the W4 SB \citep{normandeau1996galactic, dennison1997detection} and the Ophiuchus SB \citep{Pidopryhora2007} were revealed only when \HI\ and \Ha\ observations were combined.

At a distance of 4.3~kpc, considerable optical extinction is expected and it may be difficult to identify \Ha emission associated with the SB. However, if the SB has a cap, we would expect to find it in Galactic latitudes greater than $b=5^\circ$. For these latitudes, within about 1~kpc along the LOS we are looking above $z=100$~pc, the height below which most interstellar dust is concentrated \citep{guo2025vertical}. Thus, for the higher latitude regions of our data, the majority of extinction comes from the nearest 1~kpc, not the full 4.3~kpc path, and we are more likely to detect \Ha emission corresponding to the SB. 

\subsubsection{A potential cap in WHAM \Ha} \label{sec:cap_WHAM}

\begin{figure*}[ht!]
\centering
    \includegraphics[width=1\textwidth]{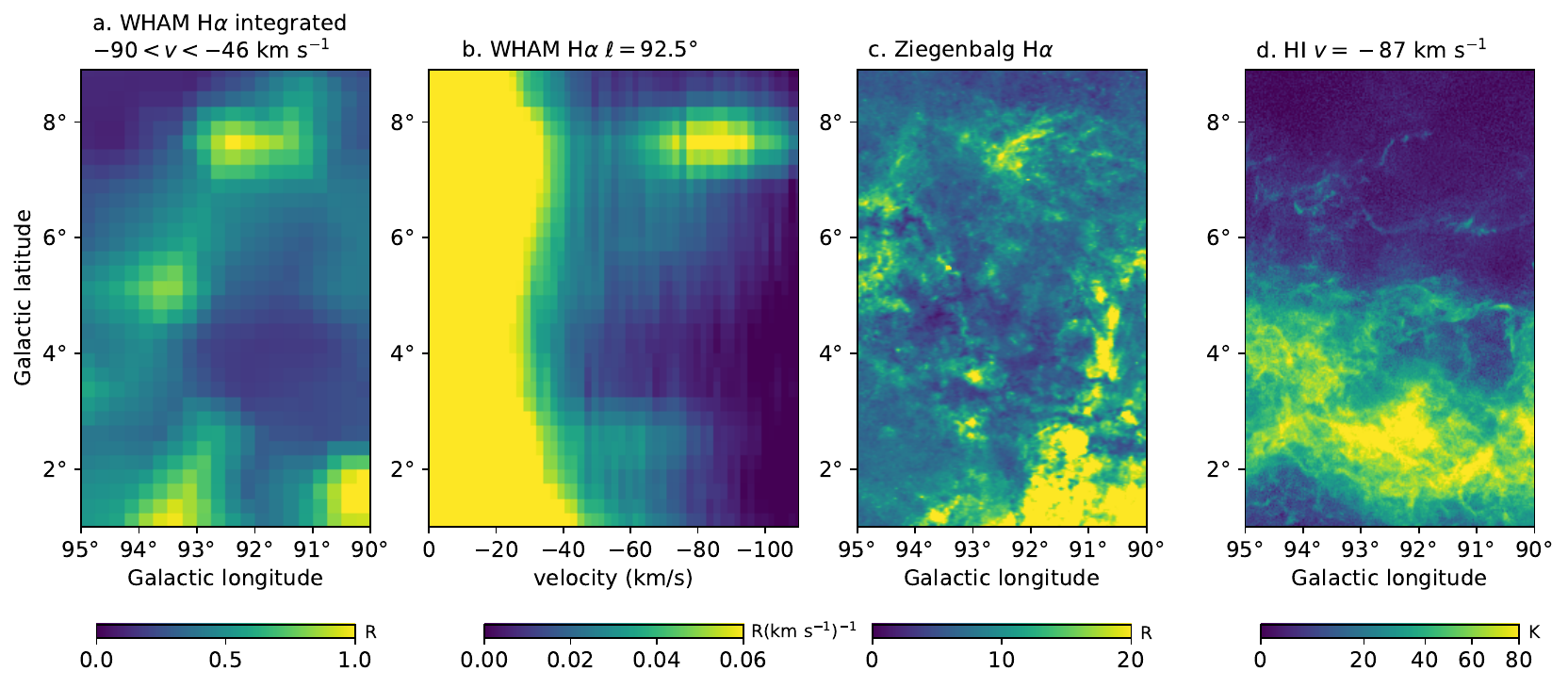}
    \caption{\textit{a.} The WHAM \Ha data cube averaged through velocities $-90 < v < -46$~\kms reveals a potential superbubble cap at latitude $b=7.7^\circ$. \textit{b.} A velocity--latitude plot of the WHAM \Ha data cube through longitude $\ell = 92.5^\circ$ shows the potential cap peaks in velocity at $v=-87$~\kms. \textit{c.} The Ziegenbalg survey \Ha map for the region shows the arched filamentary structure of the potential cap. \textit{d.} The potential superbubble cap appears in the \HI data cube at $v=-87$~\kms.}
    \label{fig:Ha1}
\end{figure*}

Figure~\ref{fig:Ha1}a shows the WHAM data cube integrated over $-90 < v < -46$\,\kms. In this image, we observe a patch of \Ha emission along latitude $b=7.7^\circ$, and spanning longitudes $91^\circ < \ell < 93^\circ$, which may be the missing cap of the SB. The $v$--$b$ diagram through $\ell = 92.5^\circ$ (Figure~\ref{fig:Ha1}b) shows that this emission peaks at $v \approx -87$\,\kms. This is  more negative than the velocity range of the potential near wall in \HI. However, such an offset would be expected if the SB expands faster into the lower-density gas at higher latitudes \citep{Kompaneets1960, basu1999dynamical}. Figure~\ref{fig:HIb}b supports this scenario, as the velocity separation of the walls appears to increase with latitude, with the near wall reaching $v \sim -85$\,\kms\ before vanishing at $b \sim 5^\circ$.

%%%%%%%%%%%%%%%%%%%%%%%% H-alpha figures %%%%%%%%%%%%%%%%%%%

\subsubsection{The potential cap in the Ziegenbalg \Ha data}

The Ziegenbalg Survey data are plotted in Figure \ref{fig:Ha1}c. Though it has been previously reported that there is no optical component to CTB~102 \citepalias{Arvidsson2009}, we note a small enhancement in \Ha emission at $(\ell, b) = (93.1^\circ, 2.7^\circ)$ corresponding with the brightest region of CTB~102 in radio continuum emission (Figure \ref{fig:continuum_and_P}). The Ziegenbalg Survey is integrated over the entire LOS, and we remain mindful that local emission is mixed with more distant emission in this dataset. Therefore, we cannot exclude the possibility that the \Ha matching the position of CTB~102 is from a more nearby emission source and is only fortuitously coincident with the \ion{H}{2} region. However, we consider it likely that some optical emission from the \HII region can overcome extinction effects and reach us. 

At higher latitudes, there is less extinction expected, and we are more likely to detect distant \Ha emission. In Figure \ref{fig:Ha1}c, we observe a curved \Ha filament centered at $(\ell, b) = (92.1^\circ, 7.7^\circ)$ that agrees with the position of the \Ha cap that we identified in the WHAM map (panel a). We conclude that they belong to the same structure because we find no emission from any other velocity range in WHAM that agrees as well in size, shape, and position. With the higher resolution of the Ziegenbalg survey, we can see that the shape of the filament agrees with what we would expect for the cap of a SB.

\subsubsection{The potential cap in \HI}

With clues about where to look for a SB cap from the \Ha data, we returned to our \HI data to look for signs of a SB closure. Figure \ref{fig:Ha1}d shows the \HI data cube at $v=-87$~\kms, the velocity where the \Ha emission from the potential SB cap peaks in WHAM. At $b=7.7^\circ$ we observe the same arched structure that we took to be the SB cap in the Ziegenbalg \Ha data. Observing this, we conclude that the SB has at least a partial cap and is not a fully open chimney.

%%%%%%%%%%%%%%%%%%%%%%%%%%%%%%%%%%%%%%%%%%%%%%%%%%%%%%%%%%%%%%%%%%%%%%%%%%%%%%%%%%%%

\section{Analysis} \label{sec:analysis}

\begin{figure*}[ht!]
\centering
    \includegraphics[width=1\textwidth]{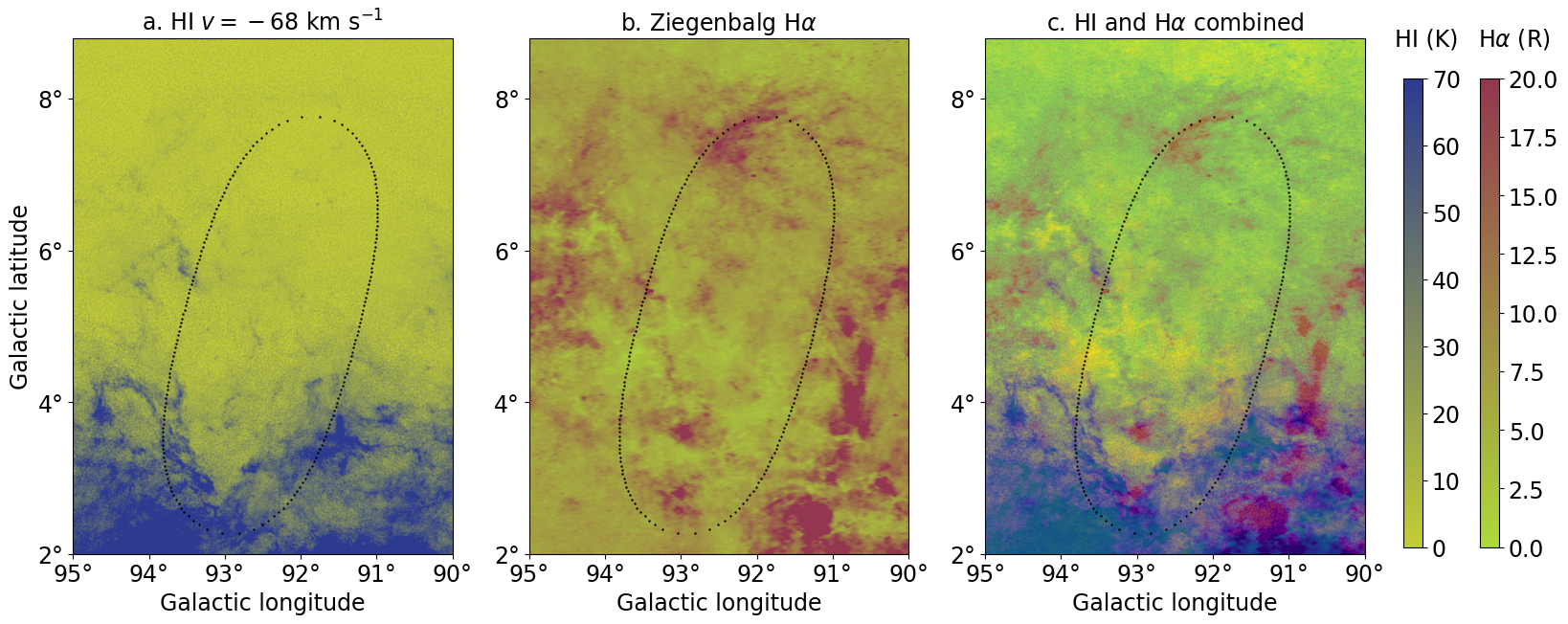}
    \caption{The outline of the \citet{Kompaneets1960} model visually fit to the CTB~102 superbubble.\textit{ a.} The model plotted on the \HI data at $v = -68$\kms. \textit{b.} The model plotted on the Ziegenbalg \Ha data. \textit{c.} The \HI and \Ha images combined.}
    \label{fig:modelfit}
\end{figure*}

We have shown that together the \HI and \Ha data present a picture of a capped SB. On Figure \ref{fig:modelfit}, we plot our estimated shape and orientation for the SB, developed by fitting the \citetalias{Kompaneets1960} to our observations (Section \ref{sec:fit_komp}). In this figure, the right \HI wall connects to a faint \Ha band at $(\ell, b) = (91.2^\circ, 4.8^\circ)$, transitioning from more neutral gas at the bottom of the SB to more ionized at the top. This apparent wall continuity supports our assertion that the \HI and \Ha emission that we have identified as SB components are part of the same structure. 

%%%%%%%%%%%%%%%%%%%%%%%%%%%%%%%%%%%%%%%%%%%%%%%%%%%%%%%%%%%%
%%%%%%%%%%%%%%%%%%% Kompaneets model on data %%%%%%%%%%%%%%%
%%%%%%%%%%%%%%%%%%%%%%%%%%%%%%%%%%%%%%%%%%%%%%%%%%%%%%%%%%%%

%%%%%%%%%%%%%%%%%% Kompaneets RM Figure %%%%%%%%%%%%%%%%%%%%
\begin{figure}[t]
\centering
    \includegraphics[width=0.47\textwidth]{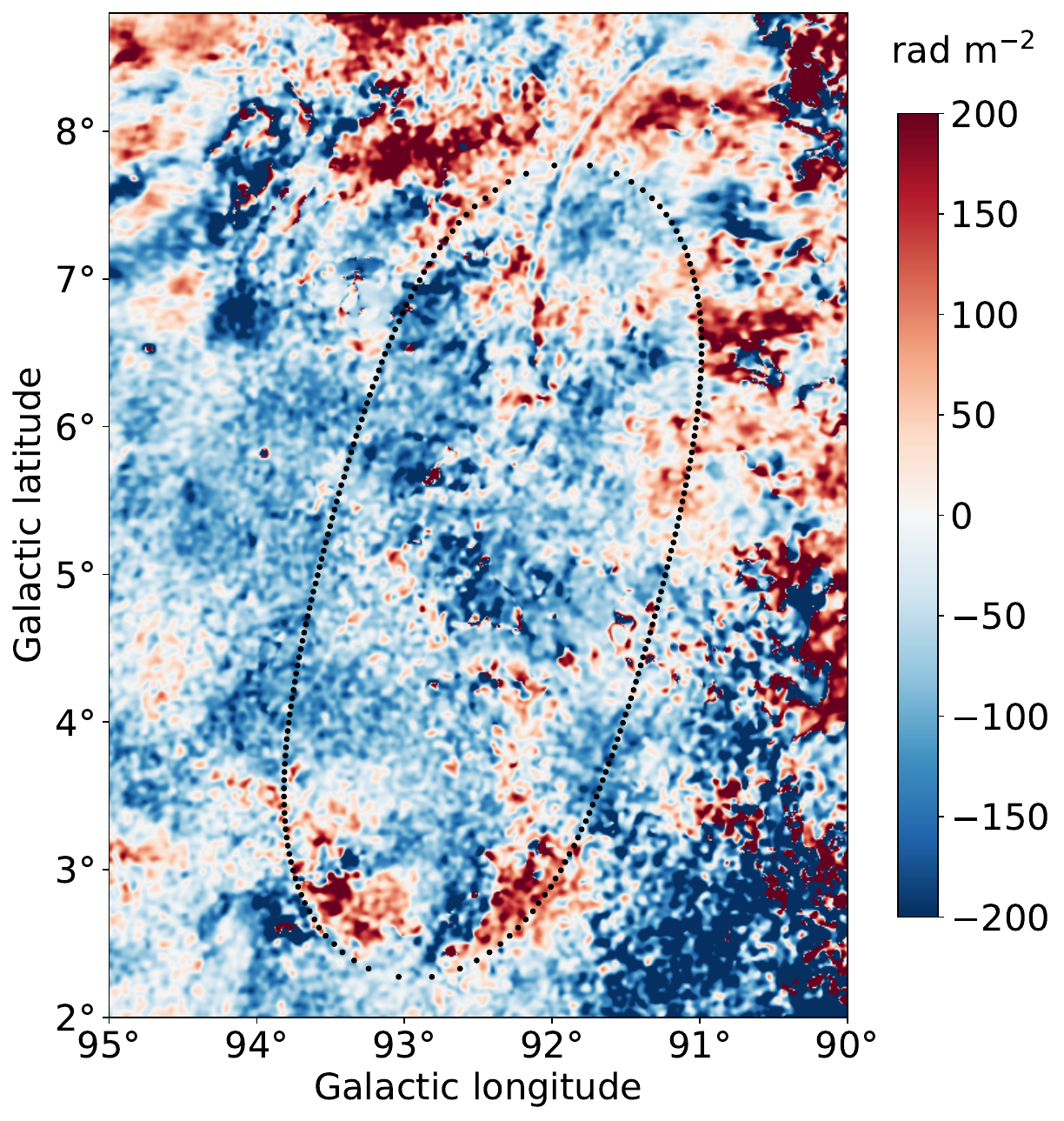}
    \caption{The \citet{Kompaneets1960} model outline from Figure \ref{fig:modelfit} plotted on the RM map of Figure \ref{fig:RMfigure}.}
    \label{fig:modelfit_RM}
\end{figure}
%%%%%%%%%%%%%%%%%%%%%%%%%%%%%%%%%%%%%%%%%%%%%%%%%%%%%%%%%%%%

The SB projection in Galactic coordinates appears as a tilted, elongated oval centered at $(\ell, b) = (92.4^\circ, 5.0^\circ)$. The SB tilts at $15^\circ$ from the vertical latitude axis and its dimensions in the tilted frame are $2.3^\circ$ wide and $5.6^\circ$ tall. At a distance of 4.3~kpc, this corresponds to minimum dimensions of $170$~pc~$\times420$~pc, which are large enough to qualify the structure as a SB. The dimensions of the SB may be greater depending on its orientation in three-dimensional space, which cannot be determined from the two-dimensional projection in Galactic coordinates. Based on the central coordinates of our modeled SB outline, we designate it as G$92.4+5.0$. However, we will continue to refer to it here as the CTB~102 SB. 

\subsection{The CTB~102 superbubble on the RM map}

In Figure \ref{fig:modelfit_RM}, we show the same CTB~102 SB outline from Figure \ref{fig:modelfit} on the RM map. In this plot, the lower boundary of the SB traces the U-shaped bowl of positive RMs. The strong spatial agreement between the SB boundary and the U-shaped RMs gives us confidence that this RM structure is related to the SB and confirms the presence of thermal electrons and magnetic fields in the lower region of the SB. We note that the top left of the SB outline traces a narrow band of positive RMs. Though the connection between this band and the SB is more tenuous, if they are associated, it would suggest similar magnetic field or electron density enhancements are present in the upper shell of the SB.

\subsection{The expansion rate of the CTB~102 superbubble} \label{sec:expand}

 In Section \ref{sec:cap_WHAM} we noted that the SB expansion rate varies with Galactic latitude, with the wall separation in velocity space greater at higher latitudes than at lower. %This is consistent with the \citetalias{Kompaneets1960}, which describes the expansion of a SB accelerating as density decreases in an exponentially stratified atmosphere. The result is a elongated bubble, in agreement with that we have observed for the CTB~102 SB.
For latitudes $b < 4^\circ$, we observe filamentary structures associated with the SB walls across the  $-74\lesssim v \lesssim -50$~\kms velocity range in our \HI data (Figures \ref{fig:HIa} and \ref{fig:HIb}). This implies a central velocity of $v\sim-62$~\kms and an expansion rate of $\sim12$~\kms. With a typical speed of sound for the warm neutral and warm ionized medium of $\sim10$~\kms \citep{hill2008turbulent}, this implies that the SB has stalled or is approaching a stalled state in the lower latitude region. In addition, we observe that the SB wall appears to be open on the left side (along $\ell \sim 94^\circ$) in panels \textit{h} to \textit{l} of Figure \ref{fig:HIa}, with filaments appearing to peel downward as if the wall is collapsing. This is a sign that, after reaching a stalled state at lower latitudes, fragmentation has begun.

At higher latitudes, there is evidence of a faster rate of expansion. At $b\sim5^\circ$ in Figure \ref{fig:HIb}b, the far and near walls vanish at $v\sim-40$~\kms and $-85$\kms, respectively, implying that the expansion rate has increased to $\sim$22.5~\kms and the SB has not yet stalled at this latitude. At $b=7.7^\circ$, the cap appears with \Ha emission peaking in the $-87$~\kms velocity channel (Figure \ref{fig:Ha1}b). As we only observe the LOS velocity component in the \Ha and \HI data, the cap of the SB may have a significant unknown velocity component perpendicular to the LOS, and the top region of the SB may be expanding more rapidly in the direction out of the Galactic disk. 

\subsection{Using the \citetalias{Kompaneets1960} model to estimate the age of the CTB~102 superbubble} \label{sec:fit_komp}

The vertical elongation of a SB has been explained using the \citetalias{Kompaneets1960} in numerous studies \citep[e.g.,][]{basu1999dynamical, Pidopryhora2007, baumgartner2013superbubble, pon2014kompaneets}. In this model, the SB is expanding in an exponentially stratified atmosphere such that density $\rho$ varies with height $z$ as
\begin{equation}
\rho = \rho_0 \, e^{-(z-z_0)/H},
\end{equation}
where $H$ is the mass-weighted scale height of all gas in the medium and $z_0$ is the position of the energy source driving the expansion. At higher elevations, the decrease in ambient density leads to an accelerated expansion rate and the SB becomes stretched out along the vertical axis. 

The radius $r$ of the SB at height $z$ is described by
\begin{equation}
r = 2H \, \arccos\big( e^{\frac{z - z_0}{2H}} \, \big[ 1 - \frac{y^2}{4H^2} + e^{-\frac{z-z_0}{H}}\big]\,\big).
\label{eq:kompR}
\end{equation}
The parameter $y$ contains information about the age of the SB through the time variable $t$, the thermal energy of the SB $E_{th}$, and the ambient density at the source of expansion $\rho_0$, as
\begin{equation}
y = \int_0^t \sqrt{ \frac{\gamma^2 - 1}{2} \, \frac{E_{th}}{\rho_0 \, \Omega(t)} } \, dt,
\label{eqn:komp_y}
\end{equation}
where $\gamma = 5/3$ is the adiabatic index for monatomic gas and $\Omega(t)$ is the volume of the SB at time $t$. The parameter $y$ is generally given in its dimensionless form, scaled by $H$ as $\tilde{y} = y/H$. The SB remains relatively spherical through $0 < \tilde{y} <1$, becomes elongated through $1 <\tilde{y} <2$, and has broken out at $\tilde{y} = 2$. To apply the \citetalias{Kompaneets1960} to observations, Equation \ref{eq:kompR} is fit to the SB in order to determine the best-fit $\tilde{y}$ and $H$. Then, Equation \ref{eqn:komp_y} is numerically evolved, incrementing $E_{th}$ and $\Omega$ with time. The time at which the integral provides the best-fit $\tilde{y}$ gives an estimated age for the SB. 

\citet{basu1999dynamical} modified the \citetalias{Kompaneets1960} to account for continuous injection of energy through stellar wind or supernovae. We fit the \citet{basu1999dynamical} formulation of the \citetalias{Kompaneets1960} to the CTB~102 SB. We started with the assumption that the stellar cluster powering CTB~102 is the energy source for the SB, and fixed the origin of the model at the position of CTB~102. The outline of the best-fit \citetalias{Kompaneets1960}, projected into Galactic coordinates, is shown in Figures \ref{fig:modelfit} and \ref{fig:modelfit_RM}. 

In order to obtain a good fit, we allowed the modeled SB to tilt in three-dimensional space such that its semi-major axis is rotated vertically through an angle $\theta$, measured from the positive $z$ direction, and horizontally through an angle $\phi$, where $\phi = 0^\circ$ away from us and increases in the counterclockwise direction as viewed from the North Galactic Pole. Including a SB tilt opens up many possible orientations for the SB in physical space, as many orientations can achieve the same projection in Galactic coordinates. In the most vertical orientation (lowest $\theta$), the best-fit values are $\theta = 15^\circ$ and $\phi = -86^\circ$, with $H =28$~pc and $\tilde{y} = 1.998$. The scale height of $H =28$~pc provides a good fit across all reasonable orientations of the SB. The parameter $\tilde{y}$ approaches 2 as the tilt of the modeled SB is increased. 

Numerically evolving Equation \ref{eqn:komp_y} according to the method described in \citet{basu1999dynamical}, we find that when $\tilde{y} = 1.998$, the age of the SB is
\begin{equation}
  t  = 6.5 \,\, \rho_0^{1/3} \, H^{5/3} \, L_0^{-1/3},
  \label{eq:komp_time}
\end{equation}
where $L_0$ is the mechanical luminosity of the stellar cluster powering the SB. Thus, to estimate $t$, we need to know the ambient density ($ \rho_0$), scale height ($H$), and mechanical luminosity. 

We use an ambient number density range between $n_0 = 0.25$~cm$^{-3}$ \citep[the ambient density around DA 530 determined by][]{Booth_2022} and 5~cm$^{-3}$ \citep[the largest reported value for the ambient density around the W4 SB as determined by][]{normandeau1996galactic}. From this range, we obtain ambient densities by assuming a mass density of 1.4 hydrogen masses per particle \citep{2005ARA&A..43..337C}. We use a mechanical luminosity range between $1\times10^{37}$~erg~$s^{-1}$ \citep[the mechanical luminosity provided by the stellar winds of a cluster containing several late-type O stars based on mass loss rates and terminal wind velocity values modeled by][]{Vink2001} and 3$\times10^{37}$ erg~$s^{-1}$ \citep[the mechanical luminosity of W4 determined by][]{normandeau1996galactic}. Using these values in Equation \ref{eq:komp_time} yields an age range between 1 and 3~Myr for the CTB~102 SB. 

\citet{stil2009three} demonstrated that ages determined using the \citetalias{Kompaneets1960} may be underestimated by as much as a factor of two due to the model neglecting the influence of magnetic fields. Thus, the age of the CTB~102 SB may be between 1~Myr and $3\times2 = 6$ Myr. Changing the orientation of the \citetalias{Kompaneets1960} only marginally increases our age estimates, as $\tilde{y} = 1.998$ represents a very evolved state in which the SB rapidly increases in height over short times. 

\section{Discussion: The origin of the CTB~102 superbubble} \label{sec:discussion}

In this section we discuss two possible scenarios for the original energy source responsible for the formation and evolution of the CTB~102 SB.

\subsection{Scenario 1: The stellar cluster powering CTB~102 is the energy source of the superbubble.} In our fitting of the \citetalias{Kompaneets1960}, we assumed that the current CTB~102 stellar cluster provided the energy that initially formed the SB and drove its expansion. This is the first origin scenario we propose for the CTB~102 SB. The best evidence for this scenario is the agreement between the \citetalias{Kompaneets1960}, using the position of CTB~102 as origin, and the observed SB morphology. A plain ellipse does not fit the observed shape of the SB as well as the \citetalias{Kompaneets1960} with CTB 102 at its origin. This is particularly true at the top and bottom of the SB where the width of an ellipse would become smaller faster than the radius described by Equation \ref{eq:kompR} for the \citetalias{Kompaneets1960}. In addition, the expansion behavior of the CTB~102 SB agrees with that predicted by the \citetalias{Kompaneets1960}, with a greater expansion velocity observed at the higher elevation regions of the SB.

The age range of 1--6~Myr determined from the \citetalias{Kompaneets1960} fitting (Section \ref{sec:fit_komp}) is also consistent with this scenario. \citetalias{Arvidsson2009} concluded that the CTB~102 \HII region is being ionized by at least several late-type O stars, which have main-sequence lifetimes of less than 10~Myr. An SB older than $\sim$10~Myr would predate these stars and rule them out as its original energy source. An upper age limit of 6~Myr makes it possible that the O stars currently powering CTB~102 were also available to drive the SB expansion throughout its lifetime. 

The significant ionization of the CTB~102 SB in the upper shell (indicated by the \Ha observations; Section \ref{sec:Ha}) and the lower shell (indicated by the RM data; Section \ref{sec:RMobs}) is evidence of the ongoing influence of the CTB~102 stellar cluster on the SB. It suggests that ionizing photons are escaping CTB~102 and reaching at least as far as the caps of the SB. This ionization explains why the upper cap is more prominent in \Ha than \HI (Figure \ref{fig:Ha1}). A parallel situation has been observed for the W4 SB, which first appeared to be an open chimney in \HI observations \citep{normandeau1996galactic}, before a closure at the top was identified in \Ha \citep{dennison1997detection}. The W4 stellar cluster is considered to be the likely source of expansion energy for the W4 SB \citep{basu1999dynamical}. The similarity between the two SBs supports a scenario in which the CTB~102 stellar cluster likewise drives the expansion of the CTB~102 SB.

\subsection{Scenario 2: The stellar cluster powering CTB~102 formed in the walls of an older superbubble.}

Our strongest arguments in support of the first scenario depend on the \citetalias{Kompaneets1960} fitting and age estimates. However, this model assumes expansion in a uniformly stratified atmosphere, inconsistent with the reality of an ISM which is structured by previous supernovae, turbulence, and other inputs. The scale height of 28 pc required to fit the model to the CTB~102 SB falls significantly short of typical scale height estimates for the ISM, which are $\sim 70$~pc for molecular gas \citep{Ferriere2001,2005ARA&A..43..337C} and $\sim 150$~pc for the cold neutral medium \citep{Kalberla2009}. This low scale height is a recurring theme that arises in attempts to fit the \citetalias{Kompaneets1960} to SB observations \citep[e.g.,][]{basu1999dynamical,Kalberla2009, baumgartner2013superbubble, pon2014kompaneets}. It is not clear that the global scale heights measured from other tracers are the scale heights experienced by SBs expanding in a local atmosphere. In the case of the W4 SB, \citet{west2007fragmenting} identified an ambient scale height of 140 pc using \HI column densities, compared to the 27~pc scale height required to fit the \citetalias{Kompaneets1960} to the SB \citep{baumgartner2013superbubble}. Given that the scale height inferred from the \citetalias{Kompaneets1960} is inconsistent with global measurements of the environment, the physical significance of the derived model parameters is unclear.

We are also uncertain of the \citetalias{Kompaneets1960} age estimate given that the wall expansion appears to have stalled, at least in the lower half of the SB (Section \ref{sec:expand}). In magnetohydrodynamic simulations of the ISM, after reaching a stalled state SBs remain as a kind of fossilized relic for many Myr before fragmenting and merging with other ISM structures \citep[e.g., see the time evolution of the simulated SB in][]{chen2026segmenting}. The \citetalias{Kompaneets1960} applied to a stalled SB would provide the age at the time of stalling, but the stalled structure may have existed for many Myr beyond that age. 

If the CTB~102 SB has stalled, it may predate the current CTB~102 stellar cluster by a large margin. Thus, we propose a second possible origin scenario: the SB originated under the power of an earlier generation of stars whose massive members are long gone, and the current CTB~102 stellar cluster formed in the expanding shock of the SB walls. This is a situation that has been identified in other SBs, such as Local Bubble \citep{Zucker2022} and the Per Tau Shell \citep{bialy2021per}, both of which have significant regions of star formation and stellar clusters observed in their walls. 

We consider both scenarios equally likely based on the evidence currently available. A reliable age determination would help to resolve this degeneracy. We would be better able to model the time evolution of the SB if we understood the surrounding magnetic environment. It has been shown that as SBs evolve, they expand more readily along the direction of the ambient magnetic field and become elongated parallel to the field lines \citep{Tomisaka1998}. \citet{west2007fragmenting} suggested that if the ambient magnetic field in the vicinity of a SB has a vertical component, then the resulting vertical elongation would mimic the elongated morphology that is characteristic of the \citetalias{Kompaneets1960}. This would explain the lower-than-expected scale height predicted by the model, and lead to a SB that is younger than the model predicts. 

Our RM map suggests that the influence of the magnetic environment in the vicinity of the SB may be substantial, as there is significant RM structure surrounding the lower shell of the SB. However, our RM map is rudimentary, having been produced by a linear fit to only four frequency channels. The renewed DRAO ST will operate from 400 to 1800 MHz with many narrow frequency channels \citep{Landecker2019}. This will enable an investigation of the three-dimensional magnetic field in the CTB~102 region using the Faraday synthesis technique \citep{brentjens2005faraday}. 

\section{The supernova remnant DA 530}

\citet{Booth_2022} determined the distance to the SNR DA 530 by observing absorption of its polarized emission by intervening \HI. Two absorptions were observed, one at $v=-28$~\kms, corresponding to a kinematic distance of 4.4 kpc, and the other at $v=-67$~\kms, corresponding to a kinematic distance of 8.3 kpc. \citet{Booth_2022} rejected the 8.3~kpc distance as it implied non-physically plausible parameters for DA 530 based on SNR expansion models. 

With the extended latitudes now available in our \HI data, we can see in the $v=-68$~\kms velocity image (Figure \ref{fig:HIa}h) that there are \HI filaments that rising up from the CTB~102 SB and crossing the region where DA~530 is located. These \HI filaments are the likely source of the $v=-67$~\kms absorption that \citet{Booth_2022} observed. This supports their rejection of the farther distance for DA~530, as these \HI filaments are associated with CTB~102, which is at 4.3~kpc \citepalias{Arvidsson2009}. 

It is of note that absorption is only observed at $v=-67$~\kms, a velocity that we have associated with the near wall of the SB (Section \ref{HIsidewalls}), and there is no absorption observed from velocities that we have associated with the far wall. Thus, it is possible that DA~530 is located within the SB. This will be another future line of investigation to be pursued with the DRAO ST following the telescope upgrade. 

\section{Summary}

We have investigated G$92.4+5.0$, a SB associated with the \HII region CTB~102. Only the lower regions of the SB (Galactic latitudes $2.5^\circ < b < 5^\circ$) are visible in \HI through velocities $-80 < v < -50$\kms. This led \citetalias{Arvidsson2009} to identify the SB as a potential chimney, and place it in the Perseus Arm at a distance of 4.3~kpc. We have identified the upper walls of the SB in \Ha, with an upper cap reaching a Galactic latitude of $b = 7.7^\circ$. Together, \HI and \Ha observations present a picture of a vertically elongated SB, tilted at an angle of $15^\circ$ from the Latitude axis in Galactic coordinates. In the tilted frame, the SB is $2.3^\circ$ wide and $5.6^\circ$ tall, which in physical dimensions approximately corresponds to 170 pc by 420 pc at a distance of 4.3~kpc. 

Detection of the upper cap of the SB was greatly aided by the high resolution provided by the Ziegenbalg \Ha data, an astrophotography project observed with private instruments. We highlight these data as a valuable tool for future investigations of the multi-phase ISM, and we would not have been able to fully understand the morphology of the SB without including observations of both the neutral and ionized gas components. 

We fit the \citetalias{Kompaneets1960} to the SB to obtain an approximate age between 1 and 6 Myr. This is consistent with the SB and CTB~102 sharing a common origin in a young stellar cluster containing O-type stars, which provide the high-energy photons that ionize both the \HII region and the walls of the SB. We also discuss an alternative scenario in which the SB was produced by an older generation of stars, and the stellar cluster currently powering CTB~102 formed in the walls of the SB. Differentiating between the two possible origin scenarios requires a reliable age estimate for the SB.

The \citetalias{Kompaneets1960} assumes the SB is expanding in an exponentially stratified medium. In the case of the CTB~102 SB, fitting the model requires a scale height of 28 pc. A smaller best-fit scale height than expected is a common outcome when fitting the \citetalias{Kompaneets1960} to SB observations. The widely discussed explanation for this discrepancy is the neglect of magnetic pressure by the model \citep{west2007fragmenting, stil2009three, pon2014kompaneets}. As this is a recurring issue, we encourage efforts to investigate the ambient magnetic fields in the vicinity of SBs and develop improved analytical models for SB evolution that incorporate both density stratification and magnetic pressure.

\section*{Acknowledgments}
This paper relies on observations obtained using a telescope at the Dominion Radio Astrophysical Observatory, which is located on the traditional, ancestral, and unceded territory of the syilx people. We benefit enormously from the stewardship of the land by the syilx Okanagan Nation and the radio frequency interference environment protection work by the syilx Okanagan Nation and NRC. DRAO is a national facility operated by the National Research Council Canada.

This research has been supported by NSERC Discovery Grants (PIs: J.B., T.L., A.S.H). R.A.B. was supported by the Natural Sciences and Engineering Research Council of Canada (NSERC) Vanier scholarship and the University of Calgary Izaak Walton Killam Doctoral Scholarship. The authors thank John Dickey for insightful discussion. 

\facilities{DRAO:Synthesis Telescope}
\software{Astropy \citep{Astropy_2022}; Matplotlib \citep{Hunter_2007}; NumPy \citep{harris2020array}; DRAO Export Package \citep{Higgs1997}}

%%%%%%%%%%%%%%%%%%%%%%%%%%%%%%%%%%%%%%%%%%%%%%%%%%%%%%%%%%%%%%%%%%%%%%%%%%%%%%%%%%%%

\bibliography{references_adstex}{}

@string{june = {June}}

@article{2005ARA&A..43..337C,
 adsurl = {https://ui.adsabs.harvard.edu/abs/2005ARA&A..43..337C},
 author = {{Cox}, Donald P.},
 doi = {10.1146/annurev.astro.43.072103.150615},
 journal = {\araa},
 month = {September},
 number = {1},
 pages = {337-385},
 title = {{The Three-Phase Interstellar Medium Revisited}},
 volume = {43},
 year = {2005}
}

@article{Arvidsson2009,
 adsurl = {https://ui.adsabs.harvard.edu/abs/2009ApJ...700.1000A},
 archiveprefix = {arXiv},
 author = {{Arvidsson}, K. and {Kerton}, C.~R. and {Foster}, T.},
 doi = {10.1088/0004-637X/700/2/1000},
 eprint = {0906.0616},
 journal = {\apj},
 month = {August},
 number = {2},
 pages = {1000-1008},
 primaryclass = {astro-ph.GA},
 title = {{The Enormous Outer Galaxy H II Region CTB 102}},
 volume = {700},
 year = {2009}
}

@article{Astropy_2022,
 adsurl = {https://ui.adsabs.harvard.edu/abs/2022ApJ...935..167A},
 archiveprefix = {arXiv},
 author = {{Astropy Collaboration} and {Price-Whelan}, Adrian M. and {Lim}, Pey Lian and {Earl}, Nicholas and {Starkman}, Nathaniel and {Bradley}, Larry and {Shupe}, David L. and {Patil}, Aarya A. and {Corrales}, Lia and {Brasseur}, C.~E. and {N{\"o}the}, Maximilian and {Donath}, Axel and {Tollerud}, Erik and {Morris}, Brett M. and {Ginsburg}, Adam and {Vaher}, Eero and {Weaver}, Benjamin A. and {Tocknell}, James and {Jamieson}, William and {van Kerkwijk}, Marten H. and {Robitaille}, Thomas P. and {Merry}, Bruce and {Bachetti}, Matteo and {G{\"u}nther}, H. Moritz and {Aldcroft}, Thomas L. and {Alvarado-Montes}, Jaime A. and {Archibald}, Anne M. and {B{\'o}di}, Attila and {Bapat}, Shreyas and {Barentsen}, Geert and {Baz{\'a}n}, Juanjo and {Biswas}, Manish and {Boquien}, M{\'e}d{\'e}ric and {Burke}, D.~J. and {Cara}, Daria and {Cara}, Mihai and {Conroy}, Kyle E. and {Conseil}, Simon and {Craig}, Matthew W. and {Cross}, Robert M. and {Cruz}, Kelle L. and {D'Eugenio}, Francesco and {Dencheva}, Nadia and {Devillepoix}, Hadrien A.~R. and {Dietrich}, J{\"o}rg P. and {Eigenbrot}, Arthur Davis and {Erben}, Thomas and {Ferreira}, Leonardo and {Foreman-Mackey}, Daniel and {Fox}, Ryan and {Freij}, Nabil and {Garg}, Suyog and {Geda}, Robel and {Glattly}, Lauren and {Gondhalekar}, Yash and {Gordon}, Karl D. and {Grant}, David and {Greenfield}, Perry and {Groener}, Austen M. and {Guest}, Steve and {Gurovich}, Sebastian and {Handberg}, Rasmus and {Hart}, Akeem and {Hatfield-Dodds}, Zac and {Homeier}, Derek and {Hosseinzadeh}, Griffin and {Jenness}, Tim and {Jones}, Craig K. and {Joseph}, Prajwel and {Kalmbach}, J. Bryce and {Karamehmetoglu}, Emir and {Ka{\l}uszy{\'n}ski}, Miko{\l}aj and {Kelley}, Michael S.~P. and {Kern}, Nicholas and {Kerzendorf}, Wolfgang E. and {Koch}, Eric W. and {Kulumani}, Shankar and {Lee}, Antony and {Ly}, Chun and {Ma}, Zhiyuan and {MacBride}, Conor and {Maljaars}, Jakob M. and {Muna}, Demitri and {Murphy}, N.~A. and {Norman}, Henrik and {O'Steen}, Richard and {Oman}, Kyle A. and {Pacifici}, Camilla and {Pascual}, Sergio and {Pascual-Granado}, J. and {Patil}, Rohit R. and {Perren}, Gabriel I. and {Pickering}, Timothy E. and {Rastogi}, Tanuj and {Roulston}, Benjamin R. and {Ryan}, Daniel F. and {Rykoff}, Eli S. and {Sabater}, Jose and {Sakurikar}, Parikshit and {Salgado}, Jes{\'u}s and {Sanghi}, Aniket and {Saunders}, Nicholas and {Savchenko}, Volodymyr and {Schwardt}, Ludwig and {Seifert-Eckert}, Michael and {Shih}, Albert Y. and {Jain}, Anany Shrey and {Shukla}, Gyanendra and {Sick}, Jonathan and {Simpson}, Chris and {Singanamalla}, Sudheesh and {Singer}, Leo P. and {Singhal}, Jaladh and {Sinha}, Manodeep and {Sip{\H{o}}cz}, Brigitta M. and {Spitler}, Lee R. and {Stansby}, David and {Streicher}, Ole and {{\v{S}}umak}, Jani and {Swinbank}, John D. and {Taranu}, Dan S. and {Tewary}, Nikita and {Tremblay}, Grant R. and {de Val-Borro}, Miguel and {Van Kooten}, Samuel J. and {Vasovi{\'c}}, Zlatan and {Verma}, Shresth and {de Miranda Cardoso}, Jos{\'e} Vin{\'\i}cius and {Williams}, Peter K.~G. and {Wilson}, Tom J. and {Winkel}, Benjamin and {Wood-Vasey}, W.~M. and {Xue}, Rui and {Yoachim}, Peter and {Zhang}, Chen and {Zonca}, Andrea and {Astropy Project Contributors}},
 doi = {10.3847/1538-4357/ac7c74},
 eid = {167},
 eprint = {2206.14220},
 journal = {\apj},
 month = {August},
 number = {2},
 pages = {167},
 primaryclass = {astro-ph.IM},
 title = {{The Astropy Project: Sustaining and Growing a Community-oriented Open-source Project and the Latest Major Release (v5.0) of the Core Package}},
 volume = {935},
 year = {2022}
}

@article{basu1999dynamical,
 adsurl = {https://ui.adsabs.harvard.edu/abs/1999ApJ...516..843B},
 archiveprefix = {arXiv},
 author = {{Basu}, Shantanu and {Johnstone}, Doug and {Martin}, P.~G.},
 doi = {10.1086/307125},
 eprint = {astro-ph/9812283},
 journal = {\apj},
 month = {May},
 number = {2},
 pages = {843-862},
 primaryclass = {astro-ph},
 title = {{Dynamical Evolution and Ionization Structure of an Expanding Superbubble: Application to W4}},
 volume = {516},
 year = {1999}
}

@article{baumgartner2013superbubble,
 adsurl = {https://ui.adsabs.harvard.edu/abs/2013A&A...557A.140B},
 archiveprefix = {arXiv},
 author = {{Baumgartner}, V. and {Breitschwerdt}, D.},
 doi = {10.1051/0004-6361/201321261},
 eid = {A140},
 eprint = {1402.0194},
 journal = {\aap},
 month = {September},
 pages = {A140},
 primaryclass = {astro-ph.GA},
 title = {{Superbubble evolution in disk galaxies. I. Study of blow-out by analytical models}},
 volume = {557},
 year = {2013}
}

@article{bekhti2016hi4pi,
 adsurl = {https://ui.adsabs.harvard.edu/abs/2016A&A...594A.116H},
 archiveprefix = {arXiv},
 author = {{HI4PI Collaboration} and {Ben Bekhti}, N. and {Fl{\"o}er}, L. and {Keller}, R. and {Kerp}, J. and {Lenz}, D. and {Winkel}, B. and {Bailin}, J. and {Calabretta}, M.~R. and {Dedes}, L. and {Ford}, H.~A. and {Gibson}, B.~K. and {Haud}, U. and {Janowiecki}, S. and {Kalberla}, P.~M.~W. and {Lockman}, F.~J. and {McClure-Griffiths}, N.~M. and {Murphy}, T. and {Nakanishi}, H. and {Pisano}, D.~J. and {Staveley-Smith}, L.},
 doi = {10.1051/0004-6361/201629178},
 eid = {A116},
 eprint = {1610.06175},
 journal = {\aap},
 month = {October},
 pages = {A116},
 primaryclass = {astro-ph.GA},
 title = {{HI4PI: A full-sky H I survey based on EBHIS and GASS}},
 volume = {594},
 year = {2016}
}

@article{bialy2021per,
 adsurl = {https://ui.adsabs.harvard.edu/abs/2021ApJ...919L...5B},
 archiveprefix = {arXiv},
 author = {{Bialy}, Shmuel and {Zucker}, Catherine and {Goodman}, Alyssa and {Foley}, Michael M. and {Alves}, Jo{\~a}o and {Semenov}, Vadim A. and {Benjamin}, Robert and {Leike}, Reimar and {En{\ss}lin}, Torsten},
 doi = {10.3847/2041-8213/ac1f95},
 eid = {L5},
 eprint = {2109.09763},
 journal = {\apjl},
 month = {September},
 number = {1},
 pages = {L5},
 primaryclass = {astro-ph.GA},
 title = {{The Per-Tau Shell: A Giant Star-forming Spherical Shell Revealed by 3D Dust Observations}},
 volume = {919},
 year = {2021}
}

@article{Booth_2022,
 adsurl = {https://ui.adsabs.harvard.edu/abs/2022ApJ...941...17B},
 archiveprefix = {arXiv},
 author = {{Booth}, Rebecca A. and {Kothes}, Roland and {Landecker}, Tom and {Brown}, Jo-Anne and {Gray}, Andrew and {Foster}, Tyler and {Greisen}, Eric},
 doi = {10.3847/1538-4357/ac9cd9},
 eid = {17},
 eprint = {2210.12207},
 journal = {\apj},
 month = {December},
 number = {1},
 pages = {17},
 primaryclass = {astro-ph.GA},
 title = {{A New Distance to the Supernova Remnant DA 530 Based on H I Absorption of Polarized Emission}},
 volume = {941},
 year = {2022}
}

@article{brentjens2005faraday,
 adsurl = {https://ui.adsabs.harvard.edu/abs/2005A&A...441.1217B},
 archiveprefix = {arXiv},
 author = {{Brentjens}, M.~A. and {de Bruyn}, A.~G.},
 doi = {10.1051/0004-6361:20052990},
 eprint = {astro-ph/0507349},
 journal = {\aap},
 month = {October},
 number = {3},
 pages = {1217-1228},
 primaryclass = {astro-ph},
 title = {{Faraday rotation measure synthesis}},
 volume = {441},
 year = {2005}
}

@article{Castor1975ApJ,
 adsurl = {https://ui.adsabs.harvard.edu/abs/1975ApJ...200L.107C},
 author = {{Castor}, J. and {McCray}, R. and {Weaver}, R.},
 doi = {10.1086/181908},
 journal = {\apjl},
 month = {September},
 pages = {L107-L110},
 title = {{Interstellar bubbles.}},
 volume = {200},
 year = {1975}
}

@article{chen2026segmenting,
 adsurl = {https://ui.adsabs.harvard.edu/abs/2026ApJ..1004...47C},
 archiveprefix = {arXiv},
 author = {{Chen}, Jing-Wen and {Hill}, Alex S. and {Ordog}, Anna and {Booth}, Rebecca A. and {Shehata}, Mohamed S.},
 doi = {10.3847/1538-4357/ae5a26},
 eid = {47},
 eprint = {2603.27741},
 journal = {\apj},
 month = {June},
 number = {1},
 pages = {47},
 primaryclass = {astro-ph.GA},
 title = {{Segmenting Superbubbles in a Simulated Multiphase Interstellar Medium Using Computer Vision}},
 volume = {1004},
 year = {2026}
}

@article{deAvillez2005,
 adsurl = {https://ui.adsabs.harvard.edu/abs/2005A&A...436..585D},
 archiveprefix = {arXiv},
 author = {{de Avillez}, M.~A. and {Breitschwerdt}, D.},
 doi = {10.1051/0004-6361:20042146},
 eprint = {astro-ph/0502327},
 journal = {\aap},
 month = {June},
 number = {2},
 pages = {585-600},
 primaryclass = {astro-ph},
 title = {{Global dynamical evolution of the ISM in star forming galaxies. I. High resolution 3D simulations: Effect of the magnetic field}},
 volume = {436},
 year = {2005}
}

@article{dennison1997detection,
 adsurl = {https://ui.adsabs.harvard.edu/abs/1997ApJ...474L..31D},
 author = {{Dennison}, Brian and {Topasna}, Gregory A. and {Simonetti}, John H.},
 doi = {10.1086/310427},
 journal = {\apjl},
 month = {January},
 number = {1},
 pages = {L31-L34},
 title = {{Detection in H{\ensuremath{\alpha}} of a Supershell Associated with W4}},
 volume = {474},
 year = {1997}
}

@article{Ferriere2001,
 adsurl = {https://ui.adsabs.harvard.edu/abs/2001RvMP...73.1031F},
 archiveprefix = {arXiv},
 author = {{Ferri{\`e}re}, Katia M.},
 doi = {10.1103/RevModPhys.73.1031},
 eprint = {astro-ph/0106359},
 journal = {Reviews of Modern Physics},
 month = {October},
 number = {4},
 pages = {1031-1066},
 primaryclass = {astro-ph},
 title = {{The interstellar environment of our galaxy}},
 volume = {73},
 year = {2001}
}

@article{guo2025vertical,
 adsurl = {https://ui.adsabs.harvard.edu/abs/2025MNRAS.543.1574G},
 archiveprefix = {arXiv},
 author = {{Guo}, H.-L. and {Chen}, B.-Q. and {Yuan}, H.-B. and {Liu}, X.-W.},
 doi = {10.1093/mnras/staf1585},
 eprint = {2509.14669},
 journal = {\mnras},
 month = {October},
 number = {2},
 pages = {1574-1583},
 primaryclass = {astro-ph.GA},
 title = {{Vertical structure of interstellar dust in the Milky Way}},
 volume = {543},
 year = {2025}
}

@article{haffner2003wisconsin,
 adsurl = {https://ui.adsabs.harvard.edu/abs/2003ApJS..149..405H},
 archiveprefix = {arXiv},
 author = {{Haffner}, L.~M. and {Reynolds}, R.~J. and {Tufte}, S.~L. and {Madsen}, G.~J. and {Jaehnig}, K.~P. and {Percival}, J.~W.},
 doi = {10.1086/378850},
 eprint = {astro-ph/0309117},
 journal = {\apjs},
 month = {December},
 number = {2},
 pages = {405-422},
 primaryclass = {astro-ph},
 title = {{The Wisconsin H{\ensuremath{\alpha}} Mapper Northern Sky Survey}},
 volume = {149},
 year = {2003}
}

@article{harris2020array,
 adsurl = {https://ui.adsabs.harvard.edu/abs/2020Natur.585..357H},
 archiveprefix = {arXiv},
 author = {{Harris}, Charles R. and {Millman}, K. Jarrod and {van der Walt}, St{\'e}fan J. and {Gommers}, Ralf and {Virtanen}, Pauli and {Cournapeau}, David and {Wieser}, Eric and {Taylor}, Julian and {Berg}, Sebastian and {Smith}, Nathaniel J. and {Kern}, Robert and {Picus}, Matti and {Hoyer}, Stephan and {van Kerkwijk}, Marten H. and {Brett}, Matthew and {Haldane}, Allan and {del R{\'\i}o}, Jaime Fern{\'a}ndez and {Wiebe}, Mark and {Peterson}, Pearu and {G{\'e}rard-Marchant}, Pierre and {Sheppard}, Kevin and {Reddy}, Tyler and {Weckesser}, Warren and {Abbasi}, Hameer and {Gohlke}, Christoph and {Oliphant}, Travis E.},
 doi = {10.1038/s41586-020-2649-2},
 eprint = {2006.10256},
 journal = {\nat},
 month = {September},
 number = {7825},
 pages = {357-362},
 primaryclass = {cs.MS},
 title = {{Array programming with NumPy}},
 volume = {585},
 year = {2020}
}

@article{harv10,
 adsurl = {https://ui.adsabs.harvard.edu/abs/2010ApJ...712.1157H},
 archiveprefix = {arXiv},
 author = {{Harvey-Smith}, L. and {Gaensler}, B.~M. and {Kothes}, R. and {Townsend}, R. and {Heald}, G.~H. and {Ng}, C.-Y. and {Green}, A.~J.},
 doi = {10.1088/0004-637X/712/2/1157},
 eprint = {1001.3462},
 journal = {\apj},
 month = {April},
 number = {2},
 pages = {1157-1165},
 primaryclass = {astro-ph.GA},
 title = {{Faraday Rotation of the Supernova Remnant G296.5+10.0: Evidence for a Magnetized Progenitor Wind}},
 volume = {712},
 year = {2010}
}

@inproceedings{Higgs1997,
 adsurl = {https://ui.adsabs.harvard.edu/abs/1997ASPC..125...58H},
 author = {{Higgs}, L.~A. and {Hoffmann}, A.~P. and {Willis}, A.~G.},
 booktitle = {Astronomical Data Analysis Software and Systems VI},
 editor = {{Hunt}, Gareth and {Payne}, Harry},
 month = {January},
 pages = {58},
 series = {Astronomical Society of the Pacific Conference Series},
 title = {{The DRAO Export Software Package}},
 volume = {125},
 year = {1997}
}

@article{hill2008turbulent,
 adsurl = {https://ui.adsabs.harvard.edu/abs/2008ApJ...686..363H},
 archiveprefix = {arXiv},
 author = {{Hill}, Alex S. and {Benjamin}, Robert A. and {Kowal}, Grzegorz and {Reynolds}, Ronald J. and {Haffner}, L. Matthew and {Lazarian}, Alex},
 doi = {10.1086/590543},
 eprint = {0805.0155},
 journal = {\apj},
 month = {October},
 number = {1},
 pages = {363-378},
 primaryclass = {astro-ph},
 title = {{The Turbulent Warm Ionized Medium: Emission Measure Distribution and MHD Simulations}},
 volume = {686},
 year = {2008}
}

@article{hill2012vertical,
 adsurl = {https://ui.adsabs.harvard.edu/abs/2012ApJ...750..104H},
 archiveprefix = {arXiv},
 author = {{Hill}, Alex S. and {Joung}, M. Ryan and {Mac Low}, Mordecai-Mark and {Benjamin}, Robert A. and {Haffner}, L. Matthew and {Klingenberg}, Christian and {Waagan}, Knut},
 doi = {10.1088/0004-637X/750/2/104},
 eid = {104},
 eprint = {1202.0552},
 journal = {\apj},
 month = {May},
 number = {2},
 pages = {104},
 primaryclass = {astro-ph.GA},
 title = {{Vertical Structure of a Supernova-driven Turbulent, Magnetized Interstellar Medium}},
 volume = {750},
 year = {2012}
}

@article{hill2018there,
 adsurl = {https://ui.adsabs.harvard.edu/abs/2018Galax...6..129H},
 archiveprefix = {arXiv},
 author = {{Hill}, Alex S.},
 doi = {10.3390/galaxies6040129},
 eid = {129},
 eprint = {1810.12008},
 journal = {Galaxies},
 month = {November},
 number = {4},
 pages = {129},
 primaryclass = {astro-ph.GA},
 title = {{Is There a Polarization Horizon?}},
 volume = {6},
 year = {2018}
}

@article{Hunter_2007,
 adsurl = {https://ui.adsabs.harvard.edu/abs/2007CSE.....9...90H},
 author = {{Hunter}, John D.},
 doi = {10.1109/MCSE.2007.55},
 journal = {Computing in Science and Engineering},
 month = {May},
 number = {3},
 pages = {90-95},
 title = {{Matplotlib: A 2D Graphics Environment}},
 volume = {9},
 year = {2007}
}

@article{Kalberla2009,
 adsurl = {https://ui.adsabs.harvard.edu/abs/2009ARA&A..47...27K},
 author = {{Kalberla}, Peter M.~W. and {Kerp}, J{\"u}rgen},
 doi = {10.1146/annurev-astro-082708-101823},
 journal = {\araa},
 month = {September},
 number = {1},
 pages = {27-61},
 title = {{The Hi Distribution of the Milky Way}},
 volume = {47},
 year = {2009}
}

@article{Kompaneets1960,
 adsurl = {https://ui.adsabs.harvard.edu/abs/1960DoSSR.130.1001K},
 author = {{Kompaneets}, D.~A.},
 journal = {Akademiia Nauk SSSR Doklady},
 month = {January},
 pages = {1001},
 volume = {130},
 year = {1960}
}

@article{landecker2000synthesis,
 adsurl = {https://ui.adsabs.harvard.edu/abs/2000A&AS..145..509L},
 archiveprefix = {arXiv},
 author = {{Landecker}, T.~L. and {Dewdney}, P.~E. and {Burgess}, T.~A. and {Gray}, A.~D. and {Higgs}, L.~A. and {Hoffmann}, A.~P. and {Hovey}, G.~J. and {Karpa}, D.~R. and {Lacey}, J.~D. and {Prowse}, N. and {Purton}, C.~R. and {Roger}, R.~S. and {Willis}, A.~G. and {Wyslouzil}, W. and {Routledge}, D. and {Vaneldik}, J.~F.},
 doi = {10.1051/aas:2000257},
 eprint = {astro-ph/0006415},
 journal = {\aaps},
 month = {September},
 pages = {509-524},
 primaryclass = {astro-ph},
 title = {{The synthesis telescope at the Dominion Radio Astrophysical Observatory}},
 volume = {145},
 year = {2000}
}

@inproceedings{Landecker2019,
 adsurl = {https://ui.adsabs.harvard.edu/abs/2019clrp.2020...37L},
 archiveprefix = {arXiv},
 author = {{Landecker}, Tom and {Belostotski}, Leo and {Brown}, Jo-Anne and {Carlson}, Brent and {Du}, Xuan and {Foster}, Tyler and {Hill}, Alex S. and {Johnson}, Thomas and {Kothes}, Roland and {Rupen}, Michael and {Safi-Harb}, Samar and {Vanderlinde}, Keith and {Veidt}, Bruce and {West}, Jennifer},
 booktitle = {Canadian Long Range Plan for Astronomy and Astrophysics White Papers},
 doi = {10.5281/zenodo.3824486},
 eid = {37},
 eprint = {1910.12002},
 month = {October},
 pages = {37},
 primaryclass = {astro-ph.IM},
 title = {{DRAO Synthesis Telescope}},
 volume = {2020},
 year = {2019}
}

@article{mac1988numerical,
 adsurl = {https://ui.adsabs.harvard.edu/abs/1988ApJ...324..776M},
 author = {{Mac Low}, Mordecai-Mark and {McCray}, Richard},
 doi = {10.1086/165936},
 journal = {\apj},
 month = {January},
 pages = {776},
 title = {{Superbubbles in Disk Galaxies}},
 volume = {324},
 year = {1988}
}

@article{marshall2019high,
 adsurl = {https://ui.adsabs.harvard.edu/abs/2019ApJ...876...45M},
 archiveprefix = {arXiv},
 author = {{Marshall}, Brandon and {Kang}, Sung-ju and {Kerton}, C.~R. and {Kim}, Youngsik and {Choi}, Minho and {Kang}, Miju},
 doi = {10.3847/1538-4357/ab1264},
 eid = {45},
 eprint = {1904.00529},
 journal = {\apj},
 month = {May},
 number = {1},
 pages = {45},
 primaryclass = {astro-ph.GA},
 title = {{High-resolution Observations of the Molecular Clouds Associated with the Huge H II Region CTB 102}},
 volume = {876},
 year = {2019}
}

@article{norman1989disk,
 adsurl = {https://ui.adsabs.harvard.edu/abs/1989ApJ...345..372N},
 author = {{Norman}, Colin A. and {Ikeuchi}, Satoru},
 doi = {10.1086/167912},
 journal = {\apj},
 month = {October},
 pages = {372},
 title = {{The Disk-Halo Interaction: Superbubbles and the Structure of the Interstellar Medium}},
 volume = {345},
 year = {1989}
}

@article{normandeau1996galactic,
 adsurl = {https://ui.adsabs.harvard.edu/abs/1996Natur.380..687N},
 author = {{Normandeau}, M. and {Taylor}, A.~R. and {Dewdney}, P.~E.},
 doi = {10.1038/380687a0},
 journal = {\nat},
 month = {April},
 number = {6576},
 pages = {687-689},
 title = {{A galactic chimney in the Perseus arm of the Milky Way}},
 volume = {380},
 year = {1996}
}

@article{Oey2005,
 adsurl = {https://ui.adsabs.harvard.edu/abs/2005AJ....129..393O},
 archiveprefix = {arXiv},
 author = {{Oey}, M.~S. and {Watson}, Alan M. and {Kern}, Katie and {Walth}, Gregory L.},
 doi = {10.1086/426333},
 eprint = {astro-ph/0501136},
 journal = {\aj},
 month = {January},
 number = {1},
 pages = {393-401},
 primaryclass = {astro-ph},
 title = {{Hierarchical Triggering of Star Formation by Superbubbles in W3/W4}},
 volume = {129},
 year = {2005}
}

@article{ordog2025pioneering,
 adsurl = {https://ui.adsabs.harvard.edu/abs/2025AJ....169..312O},
 archiveprefix = {arXiv},
 author = {{Ordog}, Anna and {Brown}, Jo-Anne C. and {Landecker}, T.~L. and {Hill}, Alex S. and {Kothes}, Roland and {West}, Jennifer L. and {Dickey}, John M. and {Haverkorn}, Marijke and {Carretti}, Ettore and {Thomson}, Alec J.~M. and {Bracco}, Andrea and {Del Rizzo}, D.~A. and {Ransom}, Ryan R. and {Reid}, Robert I.},
 doi = {10.3847/1538-3881/adc929},
 eid = {312},
 eprint = {2501.10623},
 journal = {\aj},
 month = {June},
 number = {6},
 pages = {312},
 primaryclass = {astro-ph.GA},
 title = {{A Pioneering Experiment Combining Single-antenna and Aperture-synthesis Data to Measure Faraday Rotation with GMIMS and the CGPS}},
 volume = {169},
 year = {2025}
}

@article{Pidopryhora2007,
 adsurl = {https://ui.adsabs.harvard.edu/abs/2007ApJ...656..928P},
 archiveprefix = {arXiv},
 author = {{Pidopryhora}, Yurii and {Lockman}, Felix J. and {Shields}, Joseph C.},
 doi = {10.1086/510521},
 eprint = {astro-ph/0610894},
 journal = {\apj},
 month = {February},
 number = {2},
 pages = {928-942},
 primaryclass = {astro-ph},
 title = {{The Ophiuchus Superbubble: A Gigantic Eruption from the Inner Disk of the Milky Way}},
 volume = {656},
 year = {2007}
}

@article{pon2014kompaneets,
 adsurl = {https://ui.adsabs.harvard.edu/abs/2014MNRAS.444.3657P},
 archiveprefix = {arXiv},
 author = {{Pon}, Andy and {Johnstone}, Doug and {Bally}, John and {Heiles}, Carl},
 doi = {10.1093/mnras/stu1704},
 eprint = {1408.4454},
 journal = {\mnras},
 month = {November},
 number = {4},
 pages = {3657-3669},
 primaryclass = {astro-ph.GA},
 title = {{Kompaneets model fitting of the Orion-Eridanus superbubble}},
 volume = {444},
 year = {2014}
}

@article{reic04,
 adsurl = {https://ui.adsabs.harvard.edu/abs/2014A&A...561A..55R},
 archiveprefix = {arXiv},
 author = {{Reich}, W. and {Sun}, X.~H. and {Reich}, P. and {Gao}, X.~Y. and {Xiao}, L. and {Han}, J.~L.},
 doi = {10.1051/0004-6361/201322716},
 eid = {A55},
 eprint = {1311.3767},
 journal = {\aap},
 month = {January},
 pages = {A55},
 primaryclass = {astro-ph.GA},
 title = {{A Sino-German {\ensuremath{\lambda}}6 cm polarisation survey of the Galactic plane. VIII. Small-diameter sources}},
 volume = {561},
 year = {2014}
}

@article{stil2009three,
 adsurl = {https://ui.adsabs.harvard.edu/abs/2009ApJ...701..330S},
 archiveprefix = {arXiv},
 author = {{Stil}, Jeroen and {Wityk}, Nicole and {Ouyed}, Rachid and {Taylor}, A.~R.},
 doi = {10.1088/0004-637X/701/1/330},
 eprint = {0807.0057},
 journal = {\apj},
 month = {August},
 number = {1},
 pages = {330-347},
 primaryclass = {astro-ph},
 title = {{Three-dimensional Simulations of Magnetized Superbubbles: New Insights into the Importance of MHD Effects on Observed Quantities}},
 volume = {701},
 year = {2009}
}

@article{tayl03,
 adsurl = {https://ui.adsabs.harvard.edu/abs/2003AJ....125.3145T},
 author = {{Taylor}, A.~R. and {Gibson}, S.~J. and {Peracaula}, M. and {Martin}, P.~G. and {Landecker}, T.~L. and {Brunt}, C.~M. and {Dewdney}, P.~E. and {Dougherty}, S.~M. and {Gray}, A.~D. and {Higgs}, L.~A. and {Kerton}, C.~R. and {Knee}, L.~B.~G. and {Kothes}, R. and {Purton}, C.~R. and {Uyaniker}, B. and {Wallace}, B.~J. and {Willis}, A.~G. and {Durand}, D.},
 doi = {10.1086/375301},
 journal = {\aj},
 month = {June},
 number = {6},
 pages = {3145-3164},
 title = {{The Canadian Galactic Plane Survey}},
 volume = {125},
 year = {2003}
}

@article{Tomisaka1990,
 adsurl = {https://ui.adsabs.harvard.edu/abs/1990ApJ...361L...5T},
 author = {{Tomisaka}, Kohji},
 doi = {10.1086/185814},
 journal = {\apjl},
 month = {September},
 pages = {L5},
 title = {{Blowout of Superbubble in Galactic Magnetic Field}},
 volume = {361},
 year = {1990}
}

@article{Tomisaka1998,
 adsurl = {https://ui.adsabs.harvard.edu/abs/1998MNRAS.298..797T},
 archiveprefix = {arXiv},
 author = {{Tomisaka}, Kohji},
 doi = {10.1046/j.1365-8711.1998.01654.x},
 eprint = {astro-ph/9804029},
 journal = {\mnras},
 month = {August},
 number = {3},
 pages = {797-810},
 primaryclass = {astro-ph},
 title = {{Superbubbles in magnetized interstellar media: blowout or confinement?}},
 volume = {298},
 year = {1998}
}

@article{Vink2001,
 adsurl = {https://ui.adsabs.harvard.edu/abs/2001A&A...369..574V},
 archiveprefix = {arXiv},
 author = {{Vink}, Jorick S. and {de Koter}, A. and {Lamers}, H.~J.~G.~L.~M.},
 doi = {10.1051/0004-6361:20010127},
 eprint = {astro-ph/0101509},
 journal = {\aap},
 month = {April},
 pages = {574-588},
 primaryclass = {astro-ph},
 title = {{Mass-loss predictions for O and B stars as a function of metallicity}},
 volume = {369},
 year = {2001}
}

@article{Weaver1977ApJ,
 adsurl = {https://ui.adsabs.harvard.edu/abs/1977ApJ...218..377W},
 author = {{Weaver}, R. and {McCray}, R. and {Castor}, J. and {Shapiro}, P. and {Moore}, R.},
 doi = {10.1086/155692},
 journal = {\apj},
 month = {December},
 pages = {377-395},
 title = {{Interstellar bubbles. II. Structure and evolution.}},
 volume = {218},
 year = {1977}
}

@article{west2007fragmenting,
 adsurl = {https://ui.adsabs.harvard.edu/abs/2007ApJ...656..914W},
 archiveprefix = {arXiv},
 author = {{West}, Jennifer L. and {English}, Jayanne and {Normandeau}, Magdalen and {Landecker}, T.~L.},
 doi = {10.1086/510609},
 eprint = {astro-ph/0611226},
 journal = {\apj},
 month = {February},
 number = {2},
 pages = {914-927},
 primaryclass = {astro-ph},
 title = {{The Fragmenting Superbubble Associated with the H II Region W4}},
 volume = {656},
 year = {2007}
}

@article{wilson1960survey,
 adsurl = {https://ui.adsabs.harvard.edu/abs/1960PASP...72..331W},
 author = {{Wilson}, R.~W. and {Bolton}, J.~G.},
 doi = {10.1086/127538},
 journal = {\pasp},
 month = {October},
 number = {428},
 pages = {331},
 title = {{A Survey of Galactic Radiation at 960 Mc/s}},
 volume = {72},
 year = {1960}
}

@article{wolleben2021global,
 adsurl = {https://ui.adsabs.harvard.edu/abs/2021AJ....162...35W},
 archiveprefix = {arXiv},
 author = {{Wolleben}, M. and {Landecker}, T.~L. and {Douglas}, K.~A. and {Gray}, A.~D. and {Ordog}, A. and {Dickey}, J.~M. and {Hill}, A.~S. and {Carretti}, E. and {Brown}, J.~C. and {Gaensler}, B.~M. and {Han}, J.~L. and {Haverkorn}, M. and {Kothes}, R. and {Leahy}, J.~P. and {McClure-Griffiths}, N. and {McConnell}, D. and {Reich}, W. and {Taylor}, A.~R. and {Thomson}, A.~J.~M. and {West}, J.~L.},
 doi = {10.3847/1538-3881/abf7c1},
 eid = {35},
 eprint = {2106.00945},
 journal = {\aj},
 month = {July},
 number = {1},
 pages = {35},
 primaryclass = {astro-ph.GA},
 title = {{The Global Magneto-ionic Medium Survey: A Faraday Depth Survey of the Northern Sky Covering 1280-1750 MHz}},
 volume = {162},
 year = {2021}
}

@article{Ziegenbalg2025,
 adsurl = {https://ui.adsabs.harvard.edu/abs/2025RNAAS...9..227Z},
 author = {{Ziegenbalg}, Stefan},
 doi = {10.3847/2515-5172/adfec7},
 eid = {227},
 journal = {Research Notes of the American Astronomical Society},
 month = {August},
 number = {8},
 pages = {227},
 title = {{Northern Sky Narrowband Survey DR0.2: First [O III] and [S II] Results}},
 volume = {9},
 year = {2025}
}

@article{Zucker2022,
 adsurl = {https://ui.adsabs.harvard.edu/abs/2022Natur.601..334Z},
 archiveprefix = {arXiv},
 author = {{Zucker}, Catherine and {Goodman}, Alyssa A. and {Alves}, Jo{\~a}o and {Bialy}, Shmuel and {Foley}, Michael and {Speagle}, Joshua S. and {Gro{\^I}{\texttwosuperior}schedl}, Josefa and {Finkbeiner}, Douglas P. and {Burkert}, Andreas and {Khimey}, Diana and {Swiggum}, Cameren},
 doi = {10.1038/s41586-021-04286-5},
 eprint = {2201.05124},
 journal = {\nat},
 month = {January},
 number = {7893},
 pages = {334-337},
 primaryclass = {astro-ph.GA},
 title = {{Star formation near the Sun is driven by expansion of the Local Bubble}},
 volume = {601},
 year = {2022}
}
\bibliographystyle{aasjournal}

\end{document}